\documentclass[aps,prd,reprint,superscriptaddress,preprintnumbers,%
showpacs,a4paper,nofootinbib]{revtex4-1}

\usepackage{epsfig,graphics,graphicx}
\usepackage{amsmath,amssymb,mathtools,amsfonts,gensymb}
\usepackage{bm,soul}
\usepackage[usenames,dvipsnames]{color}
\usepackage{amssymb,url,psfrag,times}
\usepackage[varg]{txfonts}
\usepackage[colorlinks, pdfborder={0 0 0}]{hyperref}
\usepackage{lineno,verbatim}
\usepackage[utf8]{inputenc}
\usepackage[normalem]{ulem}
\usepackage{xcolor}
\usepackage{graphicx}
\usepackage{hyperref}

\usepackage[caption=false]{subfig}

\usepackage[english]{babel}
\usepackage{blindtext}

\definecolor{LinkColor}{HTML}{60a567}
\definecolor{CiteColor}{HTML}{60a567}
\definecolor{UrlColor}{HTML}{60a567}
\hypersetup{linkcolor=LinkColor}
\hypersetup{citecolor=CiteColor}
\hypersetup{urlcolor=UrlColor}
\def\phenomhm{\texttt{IMRPhenomHM}}
\def\phenomd{\texttt{IMRPhenomD}}

\def\hf{\tilde{h}}
\def\h22{\hf_{22}}
\def\hlm{\hf_{lm}}

\def\A22{A_{22}}
\def\vphi22{\varphi_{22}}
\def\Alm{A_{lm}}

\def\vphilm{\varphi_{lm}}

\def\f22{f_{22}}

\begin{document}

\newcommand{\Cardiff}{School of Physics and Astronomy, Cardiff}


\title{Parameter Estimation with a spinning multi-mode waveform model: IMRPhenomHM}

\author{Chinmay Kalaghatgi}
\affiliation{\Cardiff}

\author{Mark Hannam}
\affiliation{\Cardiff}

\author{Vivien Raymond}
\affiliation{\Cardiff}


\begin{abstract}

Gravitational waves from compact binary coalescence (CBC) sources can be decomposed into spherical-harmonic
multipoles, the dominant being the quadrupole ( $(l, |m|) = (2,2)$ ) modes. The contribution of sub-dominant modes
towards total signal power increases with increasing binary mass ratio and source inclination to the detector.
It is well-known that in these cases neglecting higher modes could lead to measurement biases, but these have not yet been
quantified with a higher-mode model that includes spin effects.
In this study, we use the multi-mode aligned-spin phenomenological waveform model \texttt{IMRPhenomHM}~\cite{London2018} to
investigate the effects of including multi-mode content in estimating source parameters and contrast the results with using a quadrupole-only
model (\texttt{IMRPhenomD}). We use as sources \texttt{IMRPhenomHM} and hybrid effective-one-body--numerical-relativity
waveforms over a range of
mass-ratio and inclination combinations, and recover the parameters with \texttt{IMRPhenomHM} and \texttt{IMRPhenomD}.
These allow us to quantify the accuracy of parameter measurements using a multi-mode model, the biases incurred when using a
quadrupole-only model to recover full (multi-mode) signals, and the systematic errors in the \texttt{IMRPhenomHM} model. We see that the
parameters recovered by multi-mode templates are more precise for all non-zero inclinations as compared to quadrupole templates. For
multi-mode injections, \texttt{IMRPhenomD} recovers biased parameters for non-zero inclinations with lower likelihood while
\texttt{IMRPhenomHM}-recovered parameters are accurate for most cases, and if a bias exists, it can be explained as a combined
effect of observational priors and (in the case of hybrid-NR signals) waveform inaccuracies. However, for cases where
\texttt{IMRPhenomHM} recovers biased parameters,
the bias is always significantly smaller than the corresponding \texttt{IMRPhenomD} recovery, and we conclude that \phenomhm\ will
be sufficiently accurate to allow unbiased measurements for most GW observations.
\end{abstract}

\date{\today}

\maketitle

\section*{Introduction}
The first gravitational-wave (GW) detection, from a binary black hole (BBH) merger, was achieved on the 14$^{\text{th}}$ of September,
2015~\cite{Abbott2016} by the two Advanced LIGO (aLIGO) detectors at Hanford and Livingston~\cite{TheLIGOScientific:2014jea}.
Two observing runs have since been completed by the aLIGO detectors, and from the second half of 2017, Advanced Virgo
(AdV)~\cite{TheVirgo:2014hva} joined the GW detector network, facilitating the first three-detector observation
of a BBH source~\cite{Abbott:2017oio}. During the first two observation runs of aLIGO and AdV, a total of ten binary black hole (BBH)
mergers and one binary neutron star (BNS) merger were detected~\cite{LIGOScientific:2018mvr,Abbott2017a}.
Most signal measurements were performed using waveform models that included only the dominant quadrupole harmonic,
although one signal (GW170729) showed  evidence for a binary with unequal masses, for which models that include higher
harmonics allow for improved measurements~\cite{Chatziioannou:2019dsz}. The goal of this work is to quantify the
measurement accuracy possible with higher-multipole models.

Any GW $h(\theta, \phi, \vec{\lambda},  t)$ can be decomposed in terms of spherical harmonics  with spin-weight -2, $^{-2}Y_{lm}(\theta,\phi)$,
\begin{equation}~\label{eq:strain_decomposition}
h (\theta, \phi, \vec{\lambda}, t) = \sum_{l} \sum_{m=-l}^{m=l} \ ^{-2}Y_{lm}(\theta, \phi) h_{lm} (\vec{\lambda}, t),
\end{equation}
where $h_{lm} (\vec{\lambda}, t)$ are the GW modes, and $\vec{\lambda}$ are the intrinsic parameters of the source, i.e., the
black-hole masses and spins. For coalescing binary systems with aligned spins, or in the co-precessing frame of precessing
systems~\cite{Schmidt:2010it,OShaughnessy:2011pmr,Boyle:2011gg}, the quadrupole modes $(l, |m|) = (2,2)$ are the strongest.
Relative to the corresponding quadrupole mode, the sub-dominant modes  ($l= 3, 4, 5 \dots ; |m| \in [0, l] \forall l$) are weakest for equal-mass
systems, their strength increasing with increasing mass ratio. In addition, for a given system, as the binary's inclination to the detector
is increased from face-on $(\theta = 0^{\circ})$ to edge-on $(\theta = 90^{\circ})$, the contribution of the dominant modes decreases, as does the
overall signal power, and the relative importance of sub-dominant modes grows.

Black-hole binaries in non-eccentric orbits are characterised by the two black-hole masses, $m_1$ and $m_2$, and the black-hole spins
$\mathbf{S}_1$ and $\mathbf{S}_2$. The inspiral rate of the binary (and phasing of the GW signal, which is crucial to measuring the
properties of the binary) is effected most strongly by combinations of these parameters: the chirp mass, $\mathcal{M}_c = (m_1 m_2)^{3/5} / (m_1 + m_2)^{1/5}$ (during the inspiral), and the total mass,
$M = m_1 + m_2$ (during the merger and ringdown); the mass ratio $q = m_1/m_2$, or alternatively symmetric mass ratio $\eta = m_1 m_2/M^2$;
and the weighted sum of the spin components parallel to the orbital angular momentum $\mathbf{\hat{L}}$,
$\chi_{\rm eff} = (m_1 \chi_1 + m_2 \chi_2)/M$,
where $\chi_i = \mathbf{S}_i \cdot \mathbf{\hat{L}} / m_i^2$~\cite{Poisson:1995ef,Cutler:1994ys,Ajith:2009bn,Baird:2012cu}.
The overall strength of the GW signal scales with $M/d_L$, where $d_L$ is the
distance from the source to the detector, and is also affected by the binary inclination $\theta_{\rm JN}$, which is the angle between
the total angular momentum $\mathbf{\hat{J}}$ (or equivalently $\mathbf{\hat{L}}$ for aligned-spin binaries)
and the line of sight, $\mathbf{\hat{N}}$. If there are spin components
perpendicular to $\mathbf{\hat{L}}$, then the binary's orbital plane and spins will precess, leading to modulations of the GW signal.
In general, precession has little effect on the overall GW phase~\cite{Buonanno:2002fy,Schmidt:2012rh},
and so might not strongly effect the measurement of other parameters~\cite{Brown:2012gs,Fairhurst:2019vut}.
Precession has not yet been measured in GW observations~\cite{LIGOScientific:2018mvr}, and for these reasons we focus on
aligned-spin binaries in this study,
and expect that the general results will also hold for precessing systems.

Waveform models describing the inspiral-merger-ringdown (IMR) stages of BBH mergers are available for non-spinning, aligned-spin and precessing configurations. Many of the aligned-spin waveform models model only the dominant quadrupole mode, but several multi-mode
models now exist. For non-spinning systems, there are effective-one-body (EOB)--numerical-relativity (NR) multi-mode models, \texttt{EOBNRv2HM}~\cite{PhysRevD.84.124052} and \texttt{TEOBiResumMultipoles}~\cite{Nagar:2019wds}, and ``Phenom'' models~\cite{Mehta:2017jpq,Mehta:2019wxm}. The EOBNR
model was recently extended to aligned-spin systems (\texttt{SEOBNRv4HM})~\cite{PhysRevD.98.084028}, and models
the ($l, |m|$) = (2,2), (2,1), (3,3), (4,4) and (5,5) modes. In this study we will use the phenomenological aligned-spin multi-mode
model \phenomhm~\cite{London2018}, which models the $ l = (2,3,4) $ \& $|m| = (l, l-1)$ modes; \phenomhm\ is described in more detail
in Sec.~\ref{sec:models}. Higher modes are also included in a series of surrogate models constructed from NR
waveforms~\cite{Blackman:2015pia,Blackman:2017dfb,Blackman:2017pcm,Varma:2019csw}.

Previous studies have investigated the effect of employing higher-order mode models for gravitational wave searches~\cite{PhysRevD.89.102003, Bustillo:2016gid, PhysRevD.97.023004, PhysRevD.97.024016} and provided an estimate of the systematic errors that could be incurred from neglecting higher-order modes in the template waveforms~\cite{PhysRevD.93.084019, PhysRevD.96.124024, Varma:2014jxa}. In ~\cite{Littenberg2012} and ~\cite{Graff:2015bba} the authors performed a full Bayesian analysis of the effects of including and neglecting higher-order modes in template waveforms non-spinning systems.  We summarise some notable results
relevant to the current study.

In Ref.~\cite{Varma:2014jxa}, the authors used multi-mode PN-NR hybrids as signals and computed the expected statistical and systematic errors from using quadrupole-only templates to estimate source parameters over a range of total mass and mass-ratio values, for a signal sky-averaged signal-to-noise ratio (SNR) of 8. The statistical errors are estimated from the Fisher Information Matrix, which is the
noise-weighted inner product between partial derivatives of the waveform. The authors also estimate the systematic errors by
calculating the fitting factor, which is the noise-weighted inner product between the signal and model waveforms, maximised
over the model parameters.
The effective systematic error is proportional to the difference between the best fit and true parameters. In this study, the authors found that non-inclusion of the subdominant modes in templates will lead to $\sim$10\% loss in detection rate for
$q \geq$6 and $M \geq 100 M_{\odot}$ and will lead to systematic errors larger than statistical errors for $q \geq 4$ and $M \geq 150 M_{\odot}$. The results obtained from a Fisher information matrix approximation are valid for high SNR events. To study the waveform errors for low or moderate SNRs and for a realistic assessment of the model's measurement capabilities, a full Bayesian analysis is required.

In a Bayesian analysis, the physical parameters of the source are estimated by matching the detector data with model waveforms.
Given detector data
$d(t)$ and a waveform model $h(t)$, the posterior over $\vec{\lambda}$ is,
	 \begin{equation}
	 p( \vec{\lambda} | d(t), h(t, \vec{\lambda})) = \frac{p(\vec{\lambda}|h(t, \vec{\lambda})) p(d(t) | \vec{\lambda}, h(t, \vec{\lambda}))}{p(d(t) | h(t, \vec{\lambda}) )},
	 \end{equation}
	 where $\vec{\lambda}$ is the vector of intrinsic and extrinsic parameters. $p(\vec{\lambda}|h(t, \vec{\lambda}))$, $p(d(t) | \vec{\lambda}, h(t, \vec{\lambda}))$ and $p(d(t) | h(t, \vec{\lambda}) )$ are the prior, likelihood and evidence respectively, where the likelihood is,
	 \begin{equation}
	 p(d(t) | \vec{\lambda}, h(t, \vec{\lambda})) \propto e^{-\frac{1}{2}< d(t) - h(t, \vec{\lambda}) | d(t) - h(t, \vec{\lambda}) >},
	 \end{equation}
	 and the quantity $<a|b>$ gives the noise-weighted inner product between the two functions; this is the \emph{match} if
$a$ and $b$ are descriptions of waveforms from systems with the same physical parameters, optimised over a relative time and phase
 shift. See Ref.~\cite{Veitch:2014wba} for more details on the techniques and algorithms employed for GW parameter estimation.
Accuracy of the inferred parameters depends on the accuracy of the waveform model used to simulate the real signal and the noise content of the detector data.

In Ref.~\cite{Littenberg2012}, the authors injected multimode non-spinning NR waveforms in zero-noise at different mass ratios with
fixed inclination of $60\degree$ and compared the systematic and statistical errors of the posteriors recovered by nonspinning
quadrupole-only (\texttt{EOBNRv2}) and multi-mode (\texttt{EOBNRv2HM}) waveform models for non-spinning systems.
They found that up to $q=6$ and for SNRs $\leq$ 50, the systematic errors from \texttt{EOBNRv2HM} were smaller than or comparable
to the statistical errors. The fractional systematic error (defined as the ratio between systematic bias and statistical error) for the
intrinsic parameters is consistently lower for \texttt{EOBNRv2HM} than \texttt{EOBNRv2}. Also, the posteriors were recovered at an
overall higher likelihood by \texttt{EOBNRv2HM} than \texttt{EOBNRv2} (see Fig.~2 of Ref.~\cite{Littenberg2012}).

In Ref.~\cite{Graff:2015bba}, the authors performed a comprehensive study of the effects of using \texttt{EOBNRv2HM} and \texttt{EOBNRv2} templates to recover \texttt{EOBNRv2HM} signals across a range of total mass values
($50 M_{\odot} \leq M_{total}  \leq 500 M_{\odot}$) and SNRs ($6 \leq \rho \leq 18$) for $q=1.25$ and $q=4$ systems at two inclinations
$(\theta_{JN} = 0, 60\degree)$. Consistent with Ref.~\cite{Littenberg2012}, the posteriors are recovered at an overall larger total
evidence by \texttt{EOBNRv2HM} compared to \texttt{EOBNRv2} for inclined systems; see Fig.~5 of Ref.~\cite{Graff:2015bba}.
These differences increase with increasing inclination, mass ratio and total mass.
The posteriors are better constrained by the multi-mode model than a quadrupole-only model (see Fig.~7 of Ref.~\cite{Graff:2015bba})
with lower systematic bias for inclined systems. They found that the multi-mode model constrains the inclination angle better than quadrupole-only model, which in turn leads to better constraints on the distance.

In the previous studies, the authors used non-spinning multi-mode and quadrupole-only waveforms for the Bayesian analysis, and hence, were restricted in the ($m_{1}$, $m_{2}$) space for intrinsic parameters. In this study, we will use a multi-mode aligned-spin waveform template (\phenomhm) and increase the dimensionality of the problem by one, i.e., covering the
($m_{1}$, $m_{2}$ and $\chi_{\rm eff}$) space of intrinsic parameters. Of course, the extrinsic parameter space remains the same.

 One of the aims of this study is to explore the effects of using a multi-mode waveform template (\phenomhm) on inferring
source parameters from a multi-mode signal and to contrast it with a quadrupole-only model (\phenomd). For this, we perform a set
of injections at three different mass ratios and three inclinations in zero noise using the \phenomhm\ model. This allows us to
quantify parameter errors from not including sub-dominant modes in templates, and accuracy improvements when the
sub-dominant modes are included.

\phenomhm\ is an approximate model of the sub-dominant modes and does not model all of the higher harmonics
(all modes with  $l \geq 5$, all $m = 0$ modes, and the (3,1), (4,2), (4,1) modes).
The sub-dominant modes of the model are not tuned to NR
simulations and mode-mixing effects are not modelled. With that in mind, the other aim of the study is to determine the ability of
\phenomhm\ to recover parameters of real physical signals. For that, we perform the same set of injections as for \phenomhm\ injections but with multi-mode PN-NR hybrid waveforms and compare the parameters recovered by \phenomd\ and \phenomhm.
Hybridisation is required to include the contributions to the signals below the starting frequency of the NR waveforms.

\begin{figure*}

\subfloat{
\includegraphics[width=.328\textwidth]{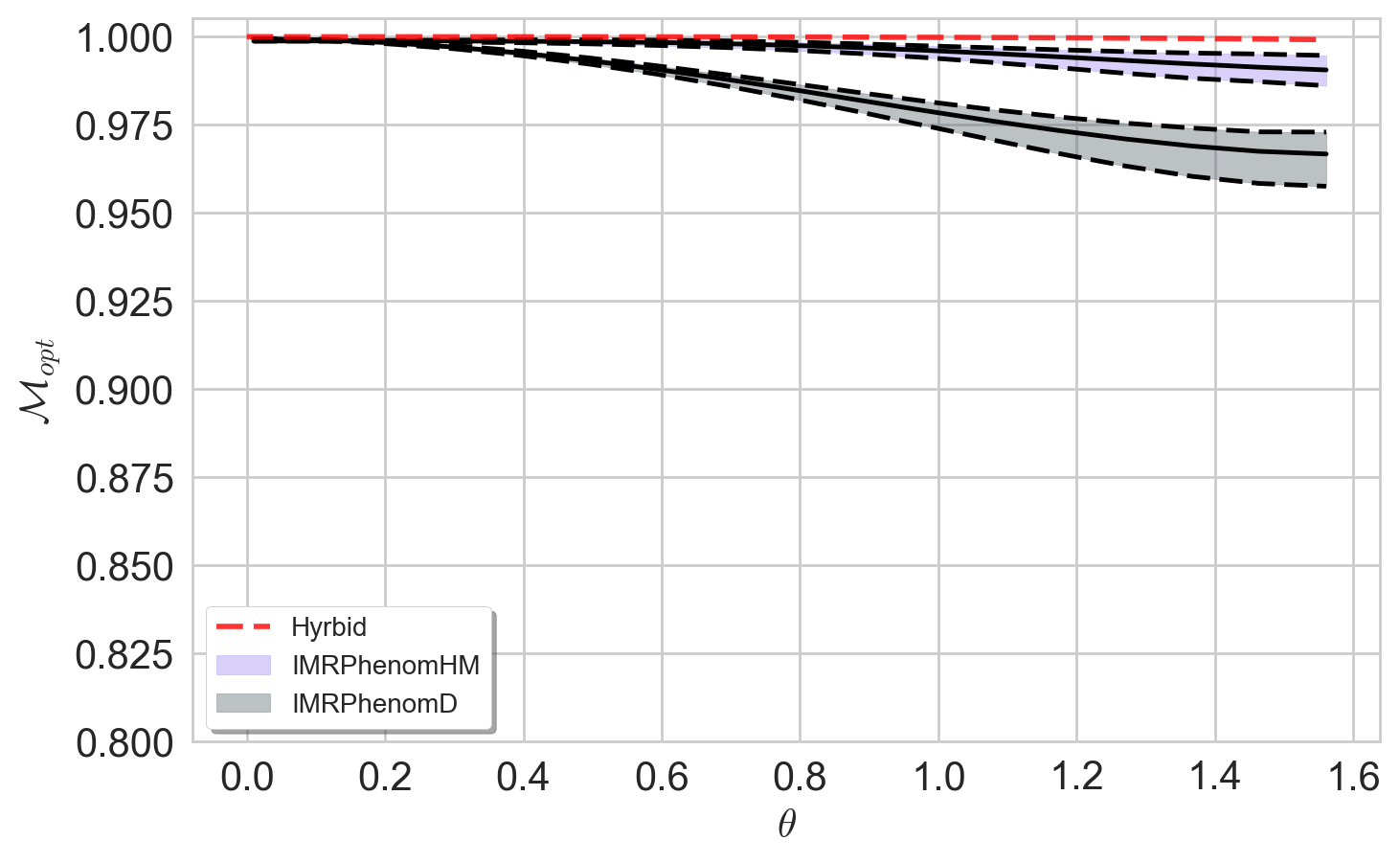}%
}
\subfloat{
\includegraphics[width=.328\textwidth]{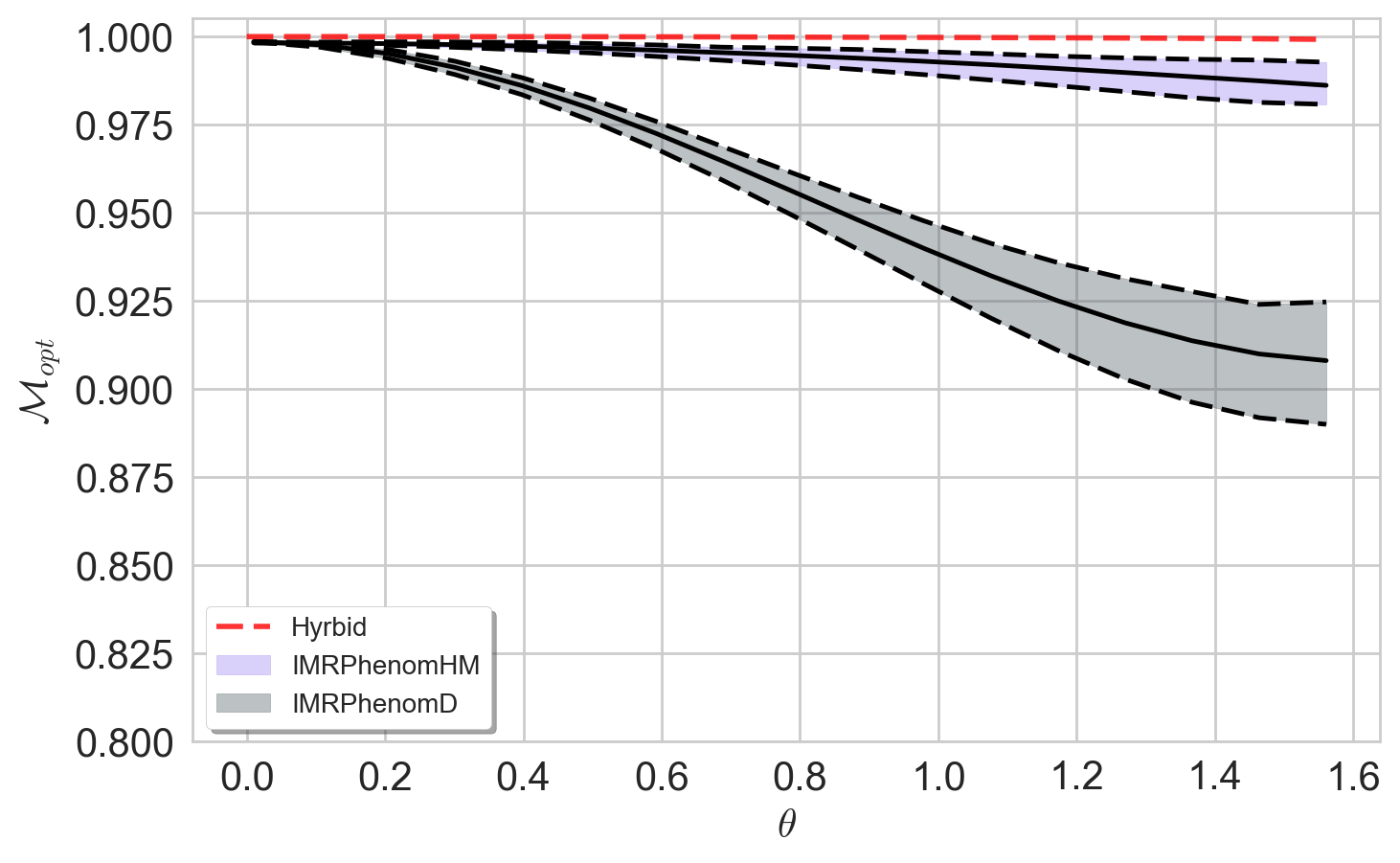}%
}
\subfloat{
\includegraphics[width=.328\textwidth]{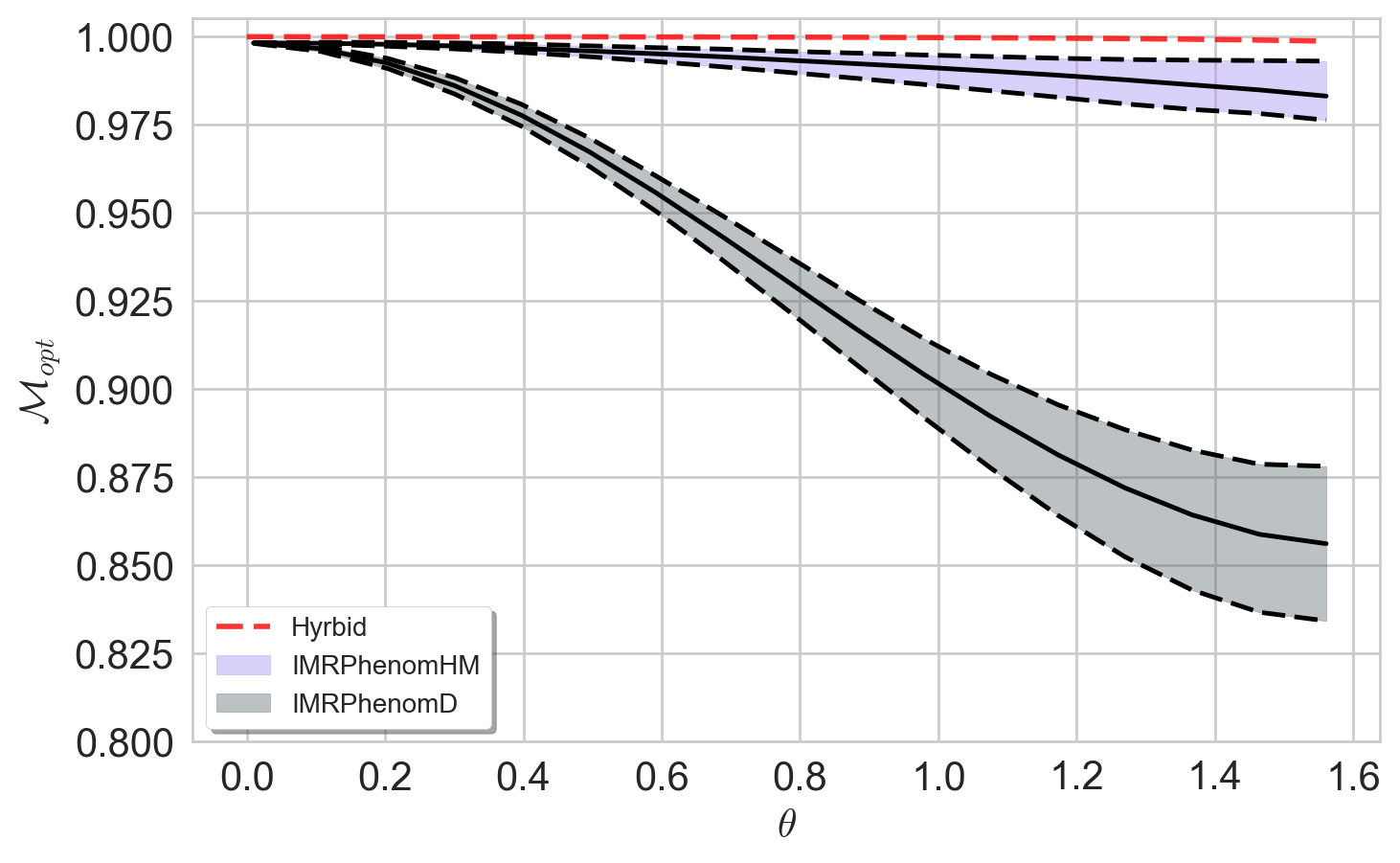}%
}
\caption{Match  between Hybrid NR and \phenomhm\ waveforms (blue) and
 \phenomd\ waveforms (grey). The left, central and right panel gives matches for $q = (2,4,8)$ systems.  The match is computed over a range of signal inclination, phase and polarisation values with the match optimized over template phase and polarisation. We quote the minimum (dashed-lower bound), average (central black line) and maximum value (dashed-upper bound) of the match at each signal inclination. The dashed red line shows the match between the multimode Hybrid-NR waveform and corresponding SXS waveform. For the match, the total mass is set to 100 M$_{\odot}$ with a lower-frequency bound of $f_{min}$ = 30Hz.}
\label{fig:nr_phnhm_matches}
\end{figure*}

	 Sec.~\ref{Methods} will provide a short summary of the template waveform models \phenomhm\ and \phenomd\, construction of the multi-mode hybrids and details of the setup for parameter estimation. The results will be introduced in general in Sec.~\ref{sec:results},
and the specific results for intrinsic and extrinsic parameters will be given in Secs.~\ref{sec:intrinsic} and \ref{sec:extrinsic}.

\section{Methods} \label{Methods}

\subsection{Summary of waveform models} \label{sec:models}

\phenomd\ \cite{Husa:2015iqa,Khan2015} is a quadrupole-only frequency-domain phenomenological waveform model describing
inspiral-merger-ringdown (IMR) stages of aligned-spin BBH systems. \phenomd\ is calibrated to NR simulations with $1 \leq q \leq 18$
and spins $\left| \chi_i \right| \lesssim 0.85$. The agreement between \phenomd\ and other quadrupole-only models is extremely good
in the region of parameter space that we consider in this study~\cite{Khan2015,Bohe:2016gbl}, and so we expect that results obtained
with this model will be indicative of the performance of any accurate quadrupole-only model.

	\phenomhm\ \cite{London2018} is a frequency domain aligned-spin phenomenological waveform model, wherein results from Post-Newtonian theory (for inspiral) and BH perturbation theory (for ringdown) are used to map the dominant quadrupole mode to the sub-dominant modes. So, given the quadrupole frequency ($f_{22}$), amplitude ($ \A22$) and phase ($\vphi22$), the frequency $(f_{lm})$, amplitude $(A_{lm})$ and phase $(\phi_{lm})$ of other modes are computed via,
	\begin{align}
	\hlm(f) &= \Alm(f) \times \exp\left\{ i\, \vphilm(f) \right\} \\
	\label{eq:MlmIntro_main}
	& \approx |\beta_{lm}(f)| \, \A22(\f22^{\rm A}) \times  \exp\left\{ i
\left[
	\kappa(f) \, \vphi22(\f22^{\rm \varphi}) + \Delta_{lm}(f)
\right] \right\} ,
\end{align}
where $ |\beta_{lm}(f)|$ and $\kappa(f)$ are amplitude and phase scaling functions and  $ \Delta_{lm}(f)$ are phase-shifts to be computed
separately for each mode.

\phenomhm\ uses  \phenomd\ for the quadrupole mode information to obtain the sub-dominant modes. Though the sub-dominant modes
of \phenomhm\ are not calibrated to NR simulations and are thus an \textit{approximation}, \phenomhm\  has much better matches to NR
waveforms than a quadrupole-only model; see Fig.~\ref{fig:nr_phnhm_matches}. We refer the reader to Ref.~\cite{London2018} for
further information on the construction and validity of the model.

\subsection{Construction and validation of multi-mode hybrids:}\
	The discussion below closely follows that of Sec.VI of Ref.~\cite{Bustillo:2015ova}.

	Two perfectly accurate gravitational waveforms [$h^{a}(t)$ \& $h^{b}(t)$] computed with different methods or with differing conventions can be mapped between each other with a time and phase shift, along with a shift in polarisation to account for differing conventions.  So, we can write $h^{a} (t)$ as,
\begin{equation}
h^{a} (t, \theta, \phi) = h^{b} (t + \tau, \theta, \phi + \phi_{0}) e^{i \psi_0} .
\end{equation}
This allows us to relate the modes of the two waveforms with each other as,
 \begin{equation}
h^{a}_{lm} (t, \theta, \phi) = h^{b}_{lm} (t + \tau, \theta) e^{i (\psi_0 + m \phi_0)} .
\end{equation}
So, given a PN inspiral waveform and a NR waveform for the same physical configuration, if the NR waveform is long enough, then there would be a common region where both the waveforms are accurate and agree with each other. We can then construct a hybrid waveform by \textit{stitching} the two within the overlapping regions. The symmetry requirements of non-precessing systems (see Eq:\ref{eq:mode_symmetry}) restrict $\psi_0$ to be either 0 or $\pi$; a more detailed discussion of which is given in Ref.~\cite{Bustillo:2015ova}.

The problem then reduces to finding appropriate time and phase shifts between the two waveforms. To obtain the time-shift ($\tau$), we minimize the quantity $\Delta (\tau, t_0, dt)$ with respect to $\tau$, where $\Delta (\tau, t_0, dt)$ is given as,
\begin{equation}
\Delta (\tau, t_0, dt) = \int_{t_0}^{t_0 + dt} \left( \omega^{PN}(t) - \omega^{NR}(t-\tau) \right)^{2}  \mathrm{d}t .
\end{equation}
Using this time shift, we construct the phase integral $\Phi(\phi_0)$,
\begin{equation}
\Phi(\phi_0) = \int_{t_0}^{t_0 + dt} \left(\phi^{NR}(t-\tau) -  \phi^{PN}(t) + \phi_{0} \right)^{2} \mathrm{d}t.
\end{equation}
The phase shift used for constructing the hybrid would be the one that minimizes $\Phi(\phi_0)$.

	Once the quantities ($\tau, \phi_0, \psi_0$) are computed, the hybrid waveform is obtained by stitching the PN and NR waveforms as,
\begin{widetext}
\begin{align}
h_{lm}(t) = \begin{cases}
		e^{i (m\phi_0 + \psi_0)} h^{PN} (t + \tau) & t < t_{0} - \tau  \\
		w^{-}(t) e^{i (m\phi_0 + \psi_0)} h^{PN} (t + \tau) + w^{+}(t) h^{NR}(t) & t_{0} - \tau < t < t_{0} - \tau + dt \\
		h^{NR}(t) & t > t_{0} - \tau + dt ,
		\end{cases}
\end{align}
\end{widetext}
where $w^{-}(t)$ \& $w^{+}(t)$ are the blending functions which go from [1,0] \& [0,1] respectively between $t_{0} - \tau < t < t_{0} - \tau + dt $. We use Planck taper windowing function~\cite{McKechan_2010} for blending the PN and NR waveforms.

To construct the hybrids for injections, we use an EOB code to generate the inspiral and hybridise it with the corresponding public SXS
NR waveform~\cite{Boyle:2019kee} following the procedure summarised above. The reason for using the SXS waveforms is that
these include more inspiral cycles than any other currently available set.
The EOB code used to obtain the inspiral modes is based on the method described in Refs.~\cite{PhysRevD.90.044018,Nagar:2017jdw}
with the fits to the parameters as published in Ref.~\cite{PhysRevD.93.044046}. Since our purpose is to construct hybrids,
we only use the inspiral contribution to these EOB waveforms.
We construct the hybrid for (2,2), (2,1), (3,3), (3,2), (3,1), (4,4), (4,3) \& (4,2) modes and the negative '$m$' modes are related to the
positive '$m$' modes by the relation,
	\begin{equation}~\label{eq:mode_symmetry}
	h_{l,m} (t, \vec{\lambda}) = (-1)^{l} h^{*}_{l, -m}(t, \vec{\lambda}).
	\end{equation}
The EOB and NR waveforms are matched at the time when the (2,2) mode has frequencies $M\omega_{2,2} = (0.072, 0.066, 0.044)$
for mass ratios $q = (2,4,8)$ with a hybridisation window of $dt = 200M$. The hybrid NR waveforms are validated by computing the match between the hyrbid waveform and
corresponding SXS waveform (which all have a starting frequency of $M\omega_{2,2} \sim 0.04$), over the $(\theta, \phi)$ space, where
the match is maximised over the phase of the hybrid waveform; see Fig.~\ref{fig:nr_phnhm_matches}.

\subsection{Setup}

For this study, we use Hybrid NR waveforms and \phenomhm\ waveforms for injections.
We create PN-NR multimode hybrids for non-spinning $q = (2, 4, 8)$ systems. The waveforms are injected at a constant SNR of 25, total mass of 100 $M_{\odot}$ and at inclinations of $0\degree$, $60\degree$ and $90\degree$.
We choose a polarisation value $(\psi)$ of 1.4 and the gps-time is set to 1186741623. For the recovery PSD, we compute the median detector PSD via BayesWave~\cite{PhysRevD.91.084034,Cornish_2015} with gps-time set to be near to the trigger-time for GW170814~\cite{Vallisneri:2014vxa}. This is so that the recovery PSD is close to the final O2 sensitivity. The accuracy of the recovered extrinsic parameters will depend strongly on the total detector response.
We choose the sky-position (for a given polarisation and gps-time) such that the total detector response for Hanford and Livingston
are of comparable value. The right ascension and declination values used are 0.2897 rad and 1.4323 rad respectively.  All waveforms are injected in zero noise. The lower frequency cut-off for both injected signal and for the parameter estimation (PE) runs is set to 20Hz. See Ref.~\cite{Schmidt:2017btt} for details regarding the frame transformations performed
during the injection and definitions of the above parameters.

	For this study we performed a total of 36 PE runs. For each of the three mass-ratio and inclination combinations, we perform
both hybrid-NR and \phenomhm\ injections.
For each injection the parameters were recovered using both \phenomd\ and \phenomhm. All signals have an SNR of 25; this is
at the high end of SNRs we would expect in aLIGO and AdV observations (at best roughly only one in fifteen observations will have
a higher SNR, assuming a detection threshold at SNR 10~\cite{Aasi:2013wya}), so provides an indication of the best measurement precision we could
achieve, as well as the worst impacts of measurement biases and systematic errors.

	The PyCBC~\cite{alex_nitz_2019_3265452} hardware injection function ($\texttt{pycbc\_generate\_hwinj}$) is used to generate the injection frames. We use the $\texttt{LALInferenceNest}$~\cite{Veitch:2014wba} pipeline to obtain the posterior samples. All runs are performed with 1024 live points.

\section{Results}
\label{sec:results}

It has already been shown in a number of previous studies~\cite{Littenberg2012,Graff:2015bba,VanDenBroeck:2006ar, OShaughnessy:2013zfw, OShaughnessy:2014shr}
that recovery using multi-mode models can improve the measurement of intrinsic and extrinsic parameters (depending on total
mass of the system). Here we present the first results that quantify these effects when recovery is performed with an aligned-spin
multi-mode IMR model. Given that there is a well-known partial degeneracy between the binary's mass ratio and the
black-hole spins~\cite{Baird:2012cu},
we expect the inclusion of spin in the recovery template to significantly affect the precision of the parameter recovery, even for signals from
non-spinning binaries. Furthermore, the effects of precession (which are not considered here), in general are driven by the in-plane spins~\cite{PhysRevLett.113.151101,Pan:2013rra} and precession measurement approximately decouples from aligned-spin parameters~\cite{Fairhurst:2019vut,Fairhurst:2019srr}  and hence, we expect that our results will in many cases carry over to
recovery using generic-binary models.

We first discuss the recovery of the intrinsic parameters (black-hole masses and spins), and then consider the extrinsic parameters, i.e.,
the distance $(d_{L})$ and inclination $(\theta_{JN})$. As we are considering only non-spinning or aligned-spin binaries,
$\vec{L} \parallel \vec{J}$ and so, $\theta_{JN} = \theta_{LN}$.

\phenomhm\ is an approximate model for the sub-dominant modes and in particular is not tuned to fully general-relativistic NR results
through the merger and ringdown. Systematic errors due to these approximations can be tested by using \phenomhm\ to recover injections
of hybrid-NR waveforms.
Also, our hybrid-NR waveforms contain extra mode-content, namely the $ (l,m) = (3,1) \ \text{and} \ (4,2)$ modes, but given that these
modes contribute less than $10 \%$ to total signal power (even for edge-on configurations), we expect that the dominant source of
systematic errors will be amplitude and phase errors in the modes that are present in the model.

The results from the \phenomhm\ injections quantify the accuracy of parameter recovery using a multi-mode model, and any biases that may be incurred
by using a quadrupole-only model. We expect these to be largely independent of the choice of model; if \phenomd\ and \phenomhm\ were
replaced by some other (accurate) quadrupole-only and multi-mode models, the results would show similar qualitative behaviours.
In contrast, the results of the hybrid-NR injections indicate the systematic errors that might be incurred when estimating the parameters of
real data using the approximate \phenomhm\ model. With that in mind, we split the results of intrinsic parameter recovered for
\phenomhm\ and hybrid-NR injections.

Our main results are in the form of the posterior distributions for each of the 18 parameter-estimation runs
(three mass ratios, three inclinations, two recovery models). The posteriors are truncated at the 90\% confidence interval, indicating the
uncertainty in each parameter measurement. Comparison against the true parameters indicates whether the measurement is unbiased
(the true value lies within the 90\% C.I.), or the level of bias. We also use the opacity of the distributions to indicate the relative log
likelihood ($\Delta \log \mathcal{L}$), which tells us how well the model agrees with the signal. In our zero-noise injections the
log likelihood is proportional to $- |h_{\rm S}(\lambda_0) - h_{\rm M}(\lambda))|$, where $h_{\rm S}$ is the injected signal, with
the specific parameters $\lambda_0$, and $h_{\rm M}$ is the template model evaluated with parameters $\lambda$, and the
magnitude of the difference between the signal and model is calculated using an inner product weighted with the detector's
spectral noise density. If the model is able to exactly reproduce the signal (as in the case of \phenomhm\ injection and recovery),
then the maximum log-likelihood will be $\log \mathcal{L} \approx 0$ and less than zero for all other cases. In our figures, an opaque
posterior distribution indicates excellent agreement between the signal and model, while a more transparent posterior distribution
indicates that the model parameters that provide the best agreement with the signal are nonetheless a poor representation of it.

Another way to quantify the difference between the performance of two models that we report is the Bayes factor, which measures the probabilistic support of one model over another, as opposed to the maximum likelihood, which provides the goodness of fit. We can
quantify the support of \phenomhm\ over \phenomd\ for any injection with the logarithm of the Bayes factor between the two models,
$\log(B^{PhnHM}_{PhnD})$. For instance, a log Bayes factor of $\log(B^{PhnHM}_{PhnD}) = 5$, means that \phenomhm\ is $e^5$
more likely than \phenomd. Note that those probabilities are given by comparing the
models \phenomhm-plus-Gaussian-noise versus \phenomd-plus-Gaussian-noise. While in this study we use zero noise as the noise
realisation, in practice Gaussian noise is only an approximation for detector noise and the Bayes factor is used with an empirically
set threshold.

\section{Recovery of intrinsic parameters}
\label{sec:intrinsic}

In this section we discuss the differences between the source mass and spin parameters ($\mathcal{M}_c , q, \chi_{\rm eff}, M_{\rm total}$)
when recovered by both \phenomd\ and \phenomhm.

Overall, we see that the \phenomd\ recovered parameters are consistently biased for inclined systems for both \phenomhm\ and hybrid-NR injections, with the bias increasing with increasing mass-ratio of the system. Parameters recovered by \phenomhm\ also show a bias for large ($q, \theta_{JN}$) hybrid-NR injections, but this bias is always smaller than the corresponding \phenomd\ recovery. If a bias exists for \phenomhm\ injection - \phenomhm\ recovery, it can be explained by marginalisation and prior effects (see Sec:~\ref{sec:phnhm_intrinsic_recov}).  %

\begin{figure*}[p]
\begin{tabular}{cc}
\includegraphics[width=0.4\textwidth]{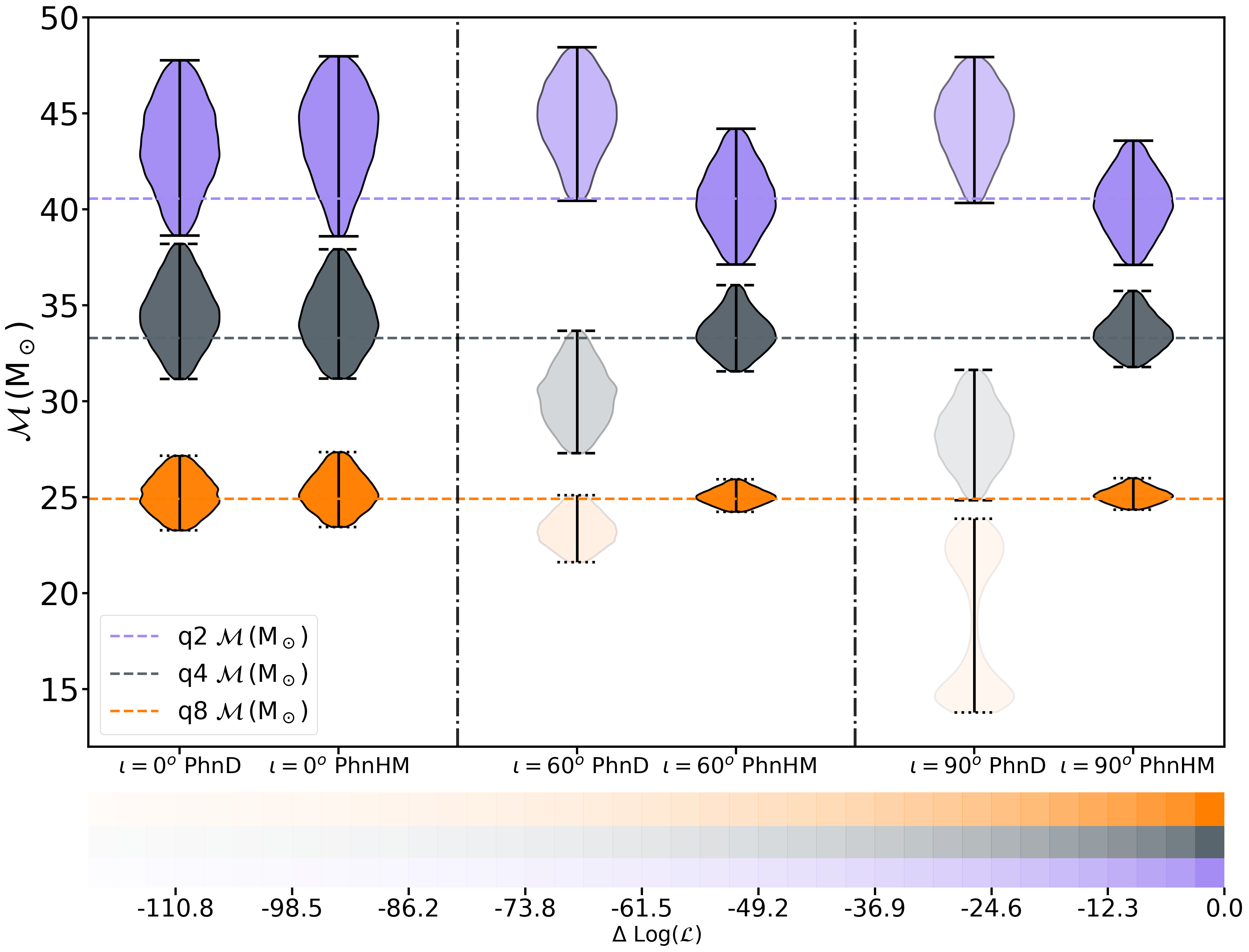}&
\includegraphics[width=0.4\textwidth]{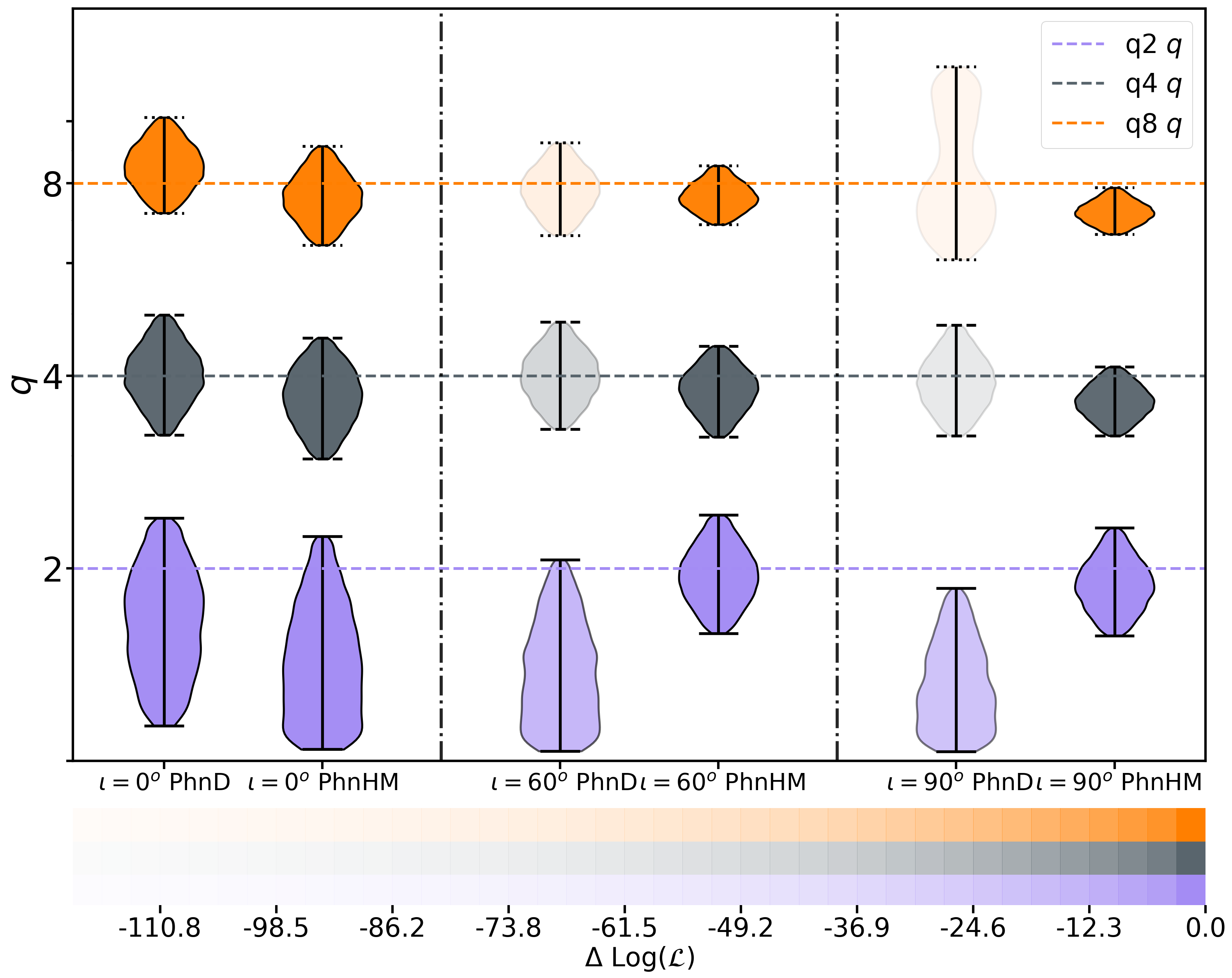} \\
\includegraphics[width=0.4\textwidth]{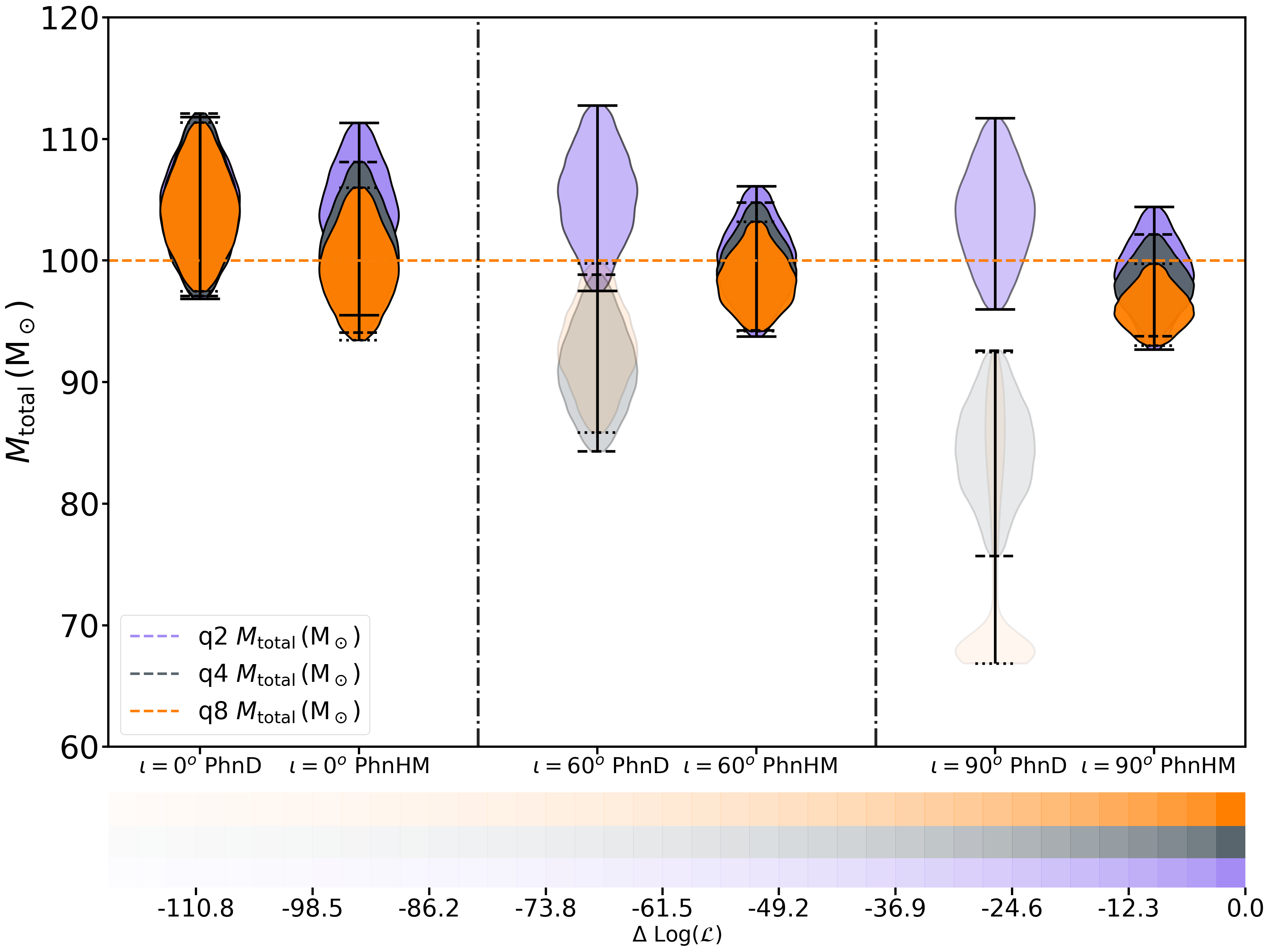}&
\includegraphics[width=0.4\textwidth]{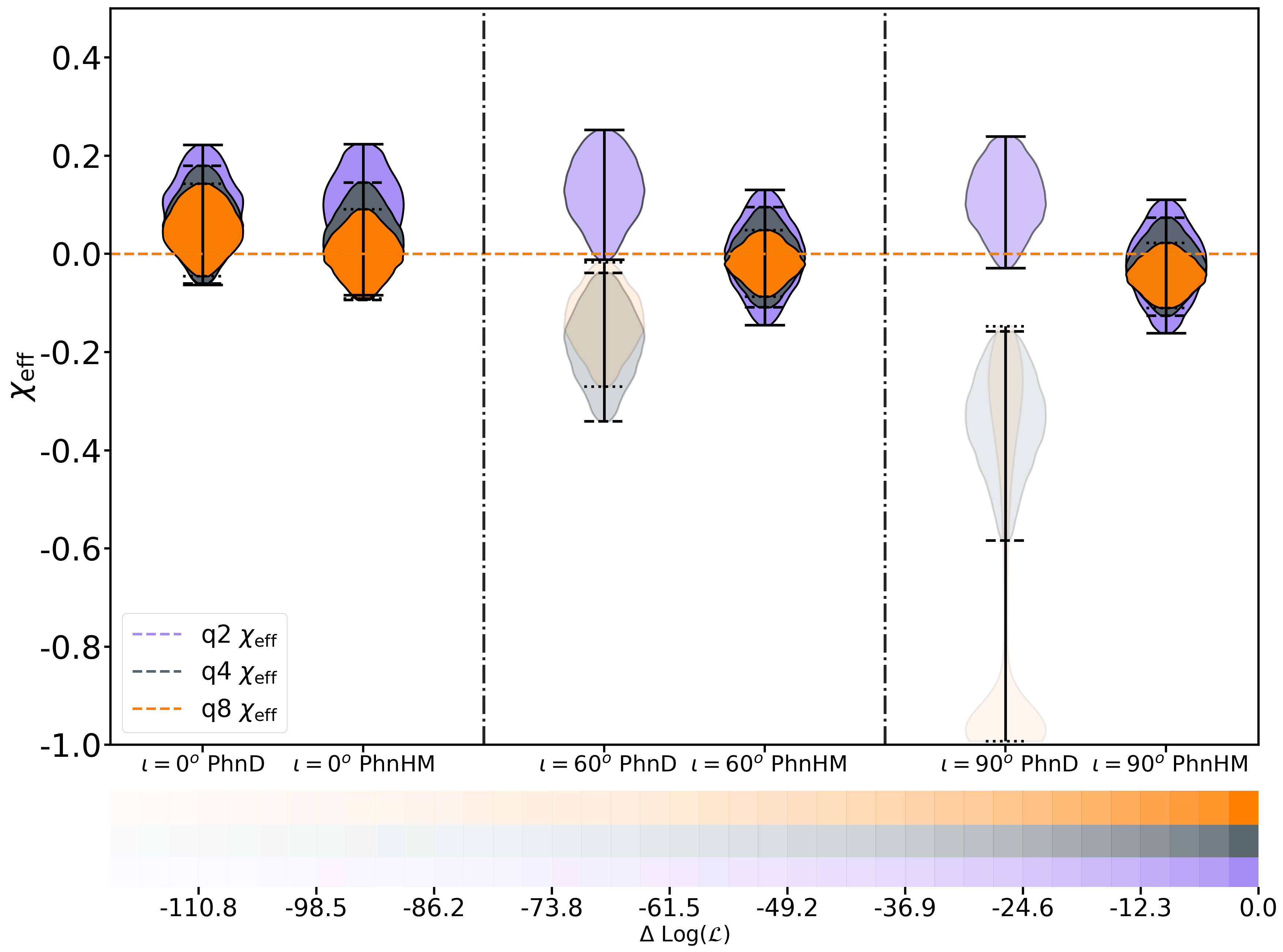} \\
\end{tabular}
\caption{Posteriors of intrinsic parameters $( \mathcal{M}_c, q, M_{\rm total} (M_{\odot}), \chi_{\rm eff})$ for \phenomhm\ waveform injected at $q = 2$, $q = 4$, $q = 8$ with $\theta_{JN} $=0, 60$^{\circ}$, 90$^{\circ}$. Posteriors for $q = 2$ ($q = 4$) [$q = 8$] are shown in Blue (Grey) [Orange] with the opacity of each determined from the maximum likelihood value of that run. The variation of opacity over the likelihood values is shown at the bottom of each graph. See Sec:~\ref{sec:phnhm_intrinsic_recov} for a description of these results, specifically, the bi-modal posteriors recovered by \phenomd\ at q=8, $\theta_{JN} = $ 90$^{o}$.
}~\label{fig:PhenomHM_intrinsic}
\vspace*{\floatsep}
\begin{tabular}{cc}
\includegraphics[width=0.4\textwidth]{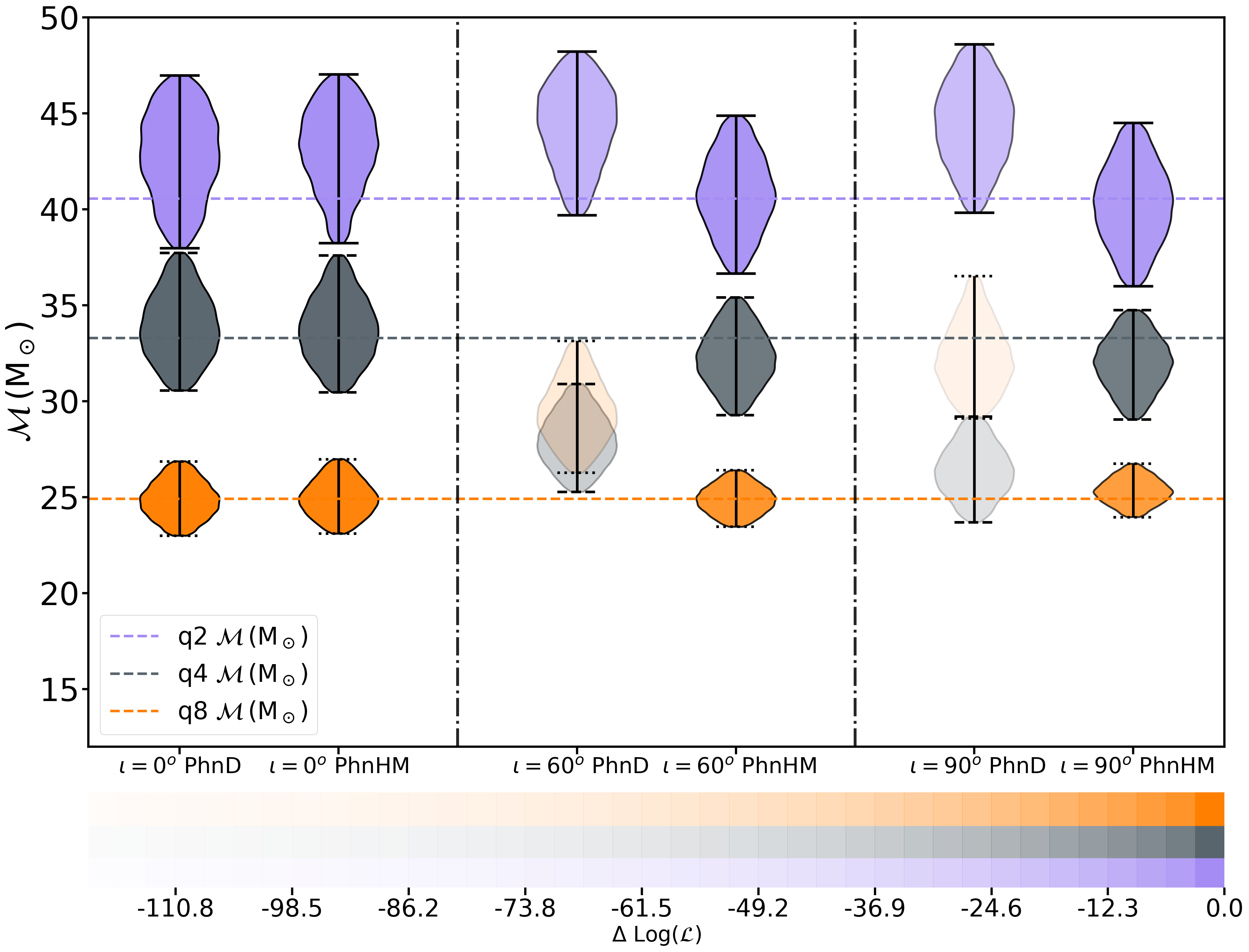}&
\includegraphics[width=0.4\textwidth]{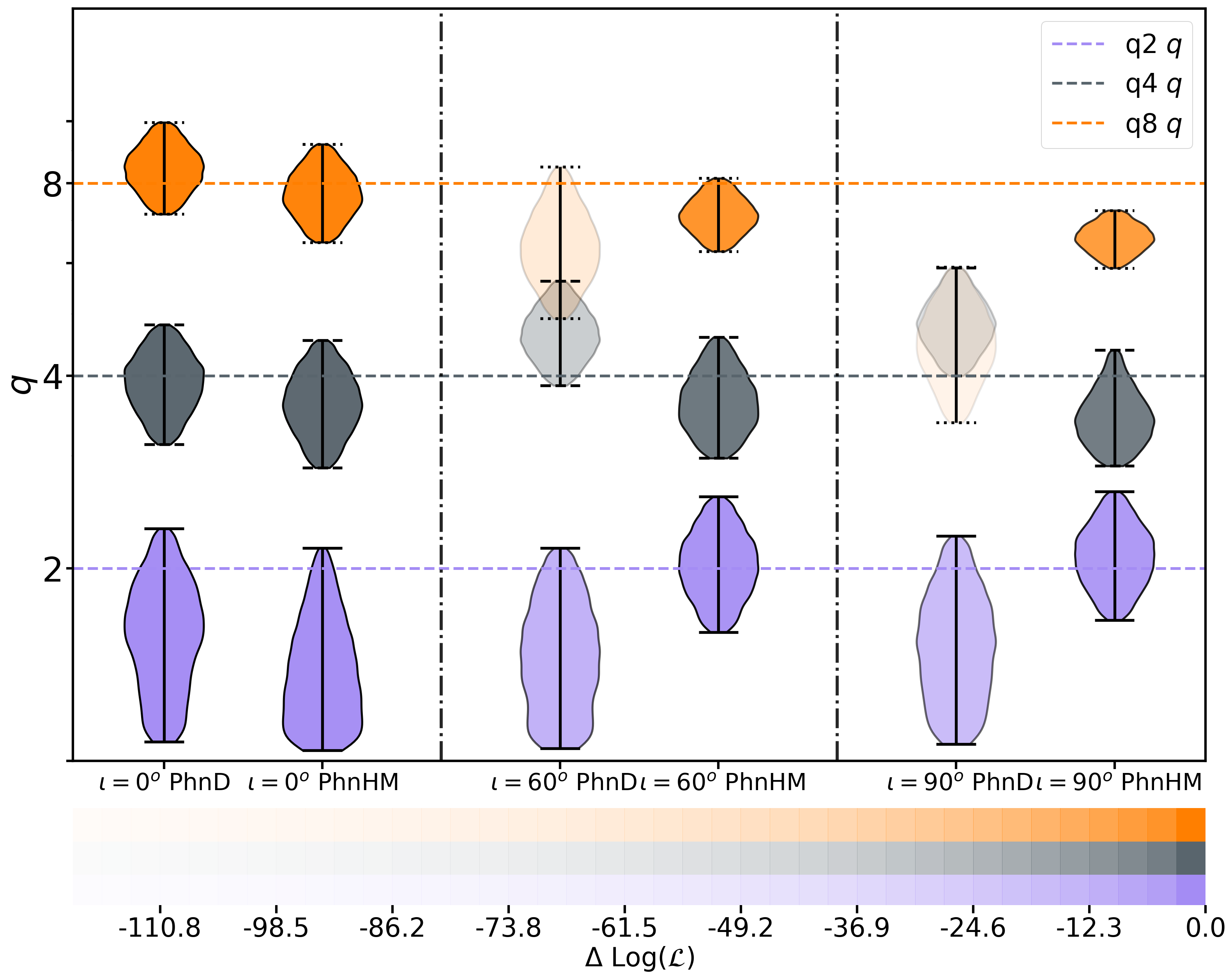} \\
\includegraphics[width=0.4\textwidth]{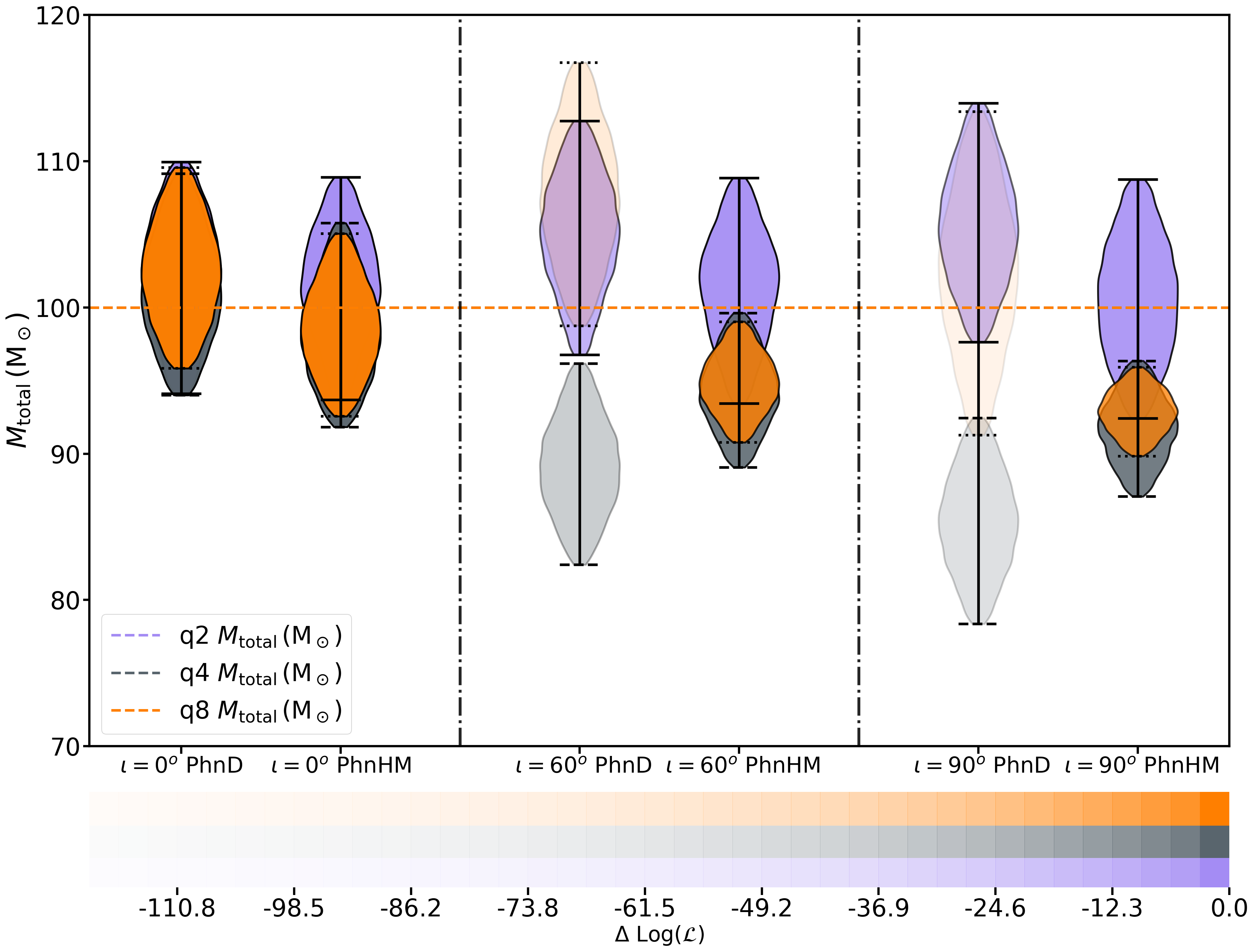}&
\includegraphics[width=0.4\textwidth]{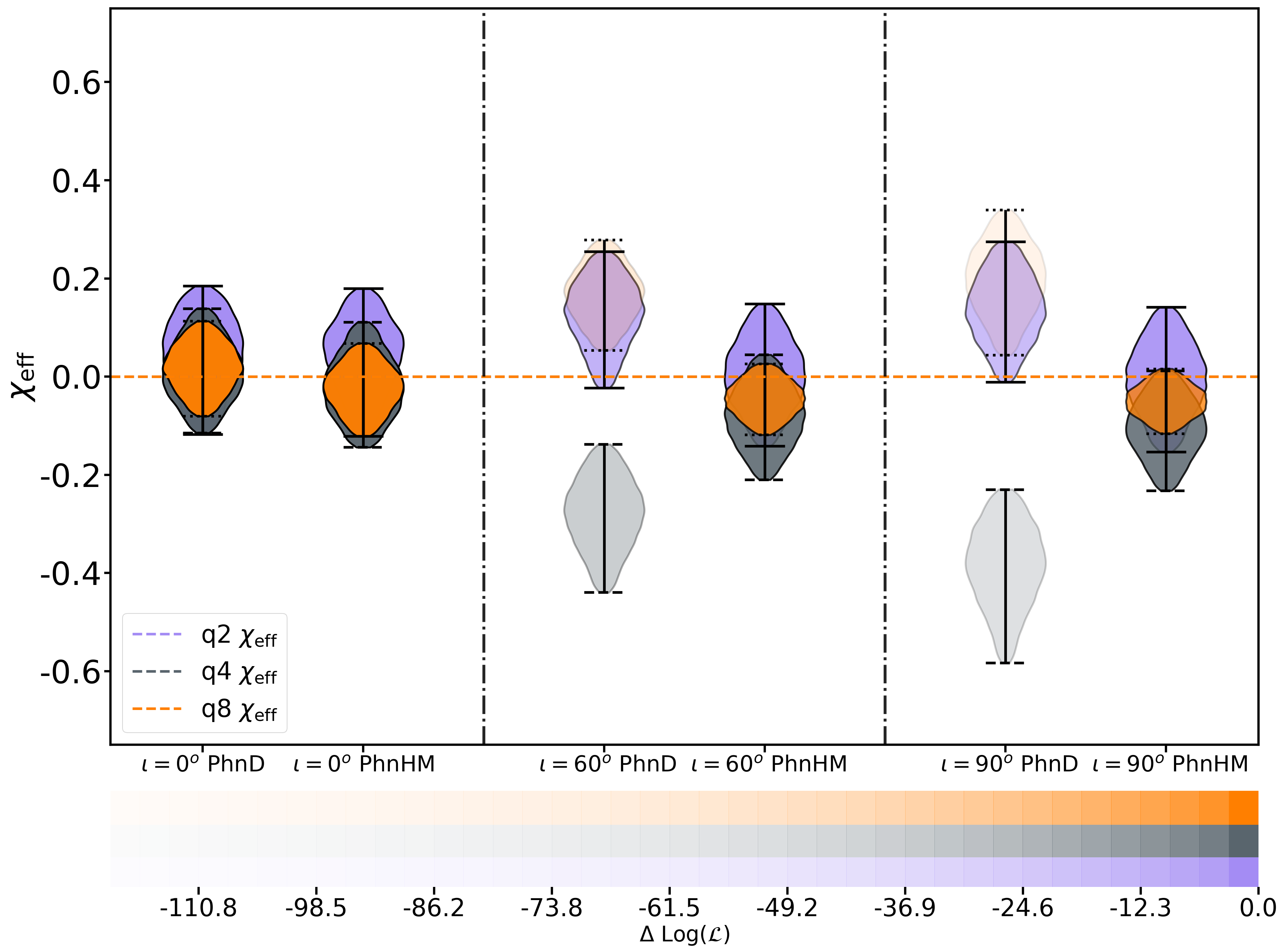} \\
\end{tabular}
\caption{Posteriors of intrinsic parameters $( \mathcal{M}_c, q, M_{\rm total}, \chi_{\rm eff})$ for Hybrid-NR waveform injected at
$q = 2$, $q = 4$, $q = 8$ with $\theta_{JN} =$ 0, 60$^{\circ}$ and 90$^{\circ}$. Posteriors for $q = 2$ ($q = 4$) [$q = 8$] are shown in Blue (Grey) [Orange] with the opacity of each determined from the maximum likelihood value of that run. The variation of opacity over the likelihood values is shown at the bottom of each graph.}~\label{fig:NR_intrinsic}

\end{figure*}

\begin{figure}
	\includegraphics[width=0.48\textwidth,height=0.5\textheight,keepaspectratio]{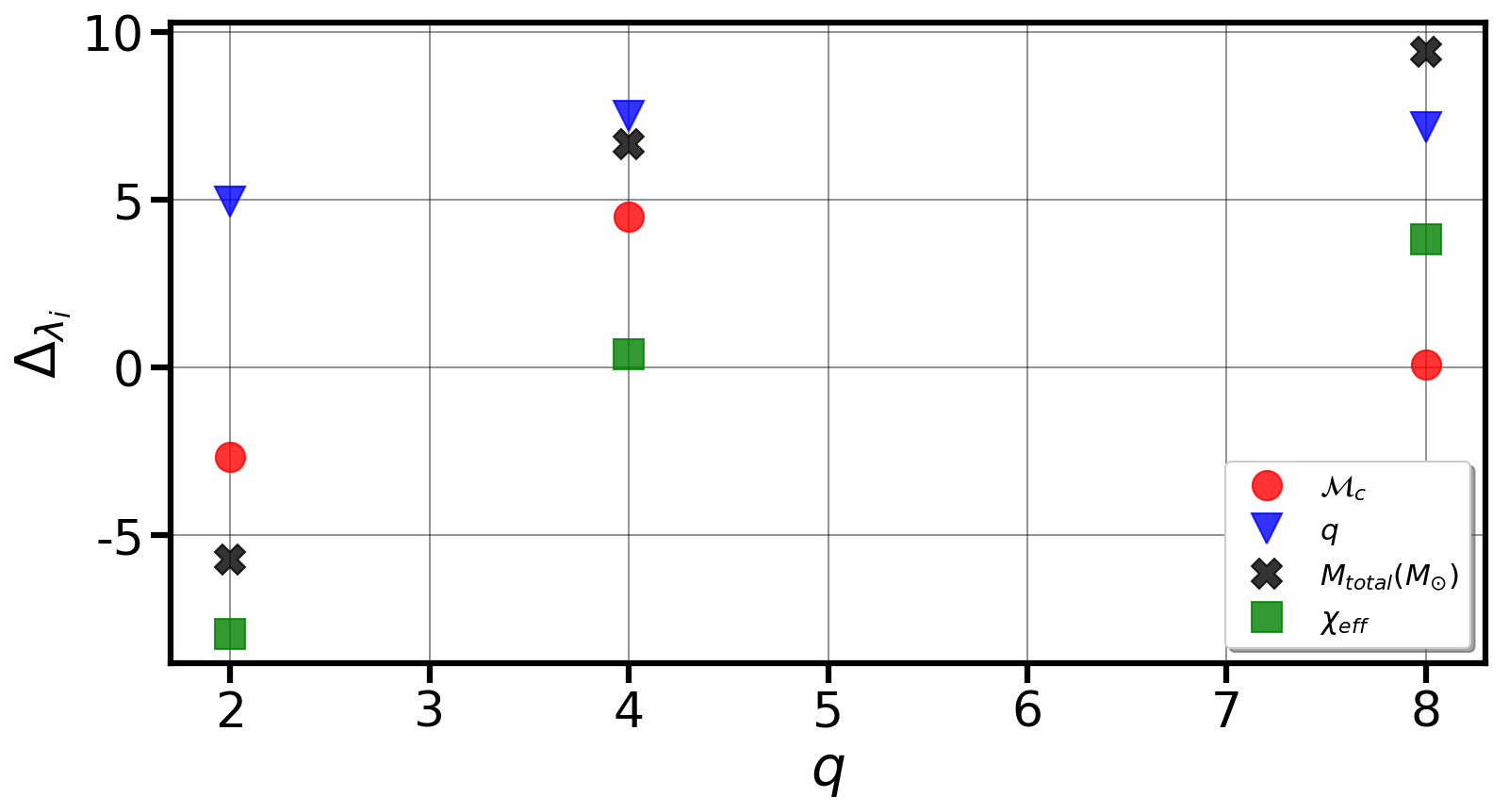}
	\includegraphics[width=0.48\textwidth,height=0.5\textheight,keepaspectratio]{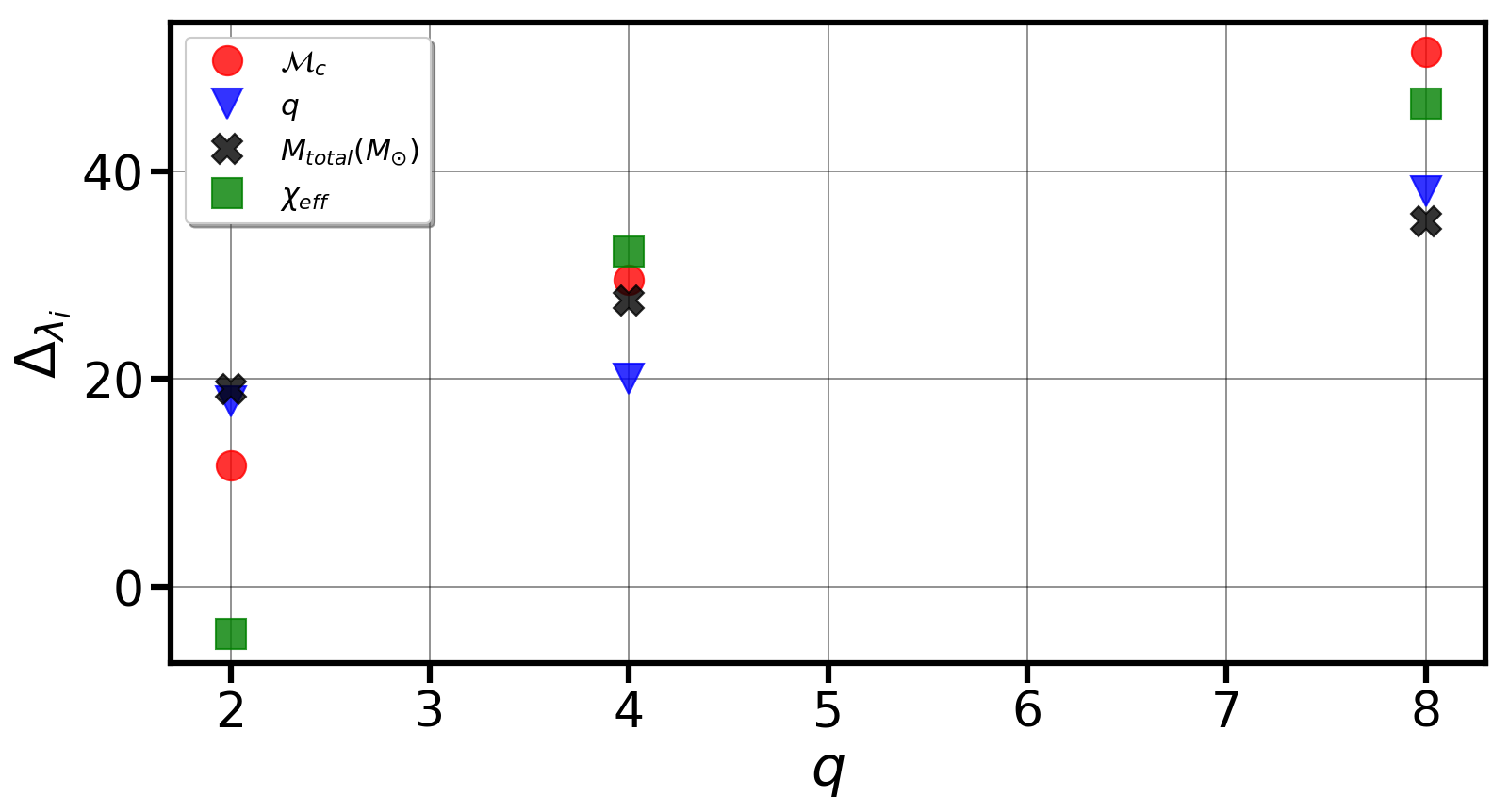}
	\includegraphics[width=0.48\textwidth,height=0.5\textheight,keepaspectratio]{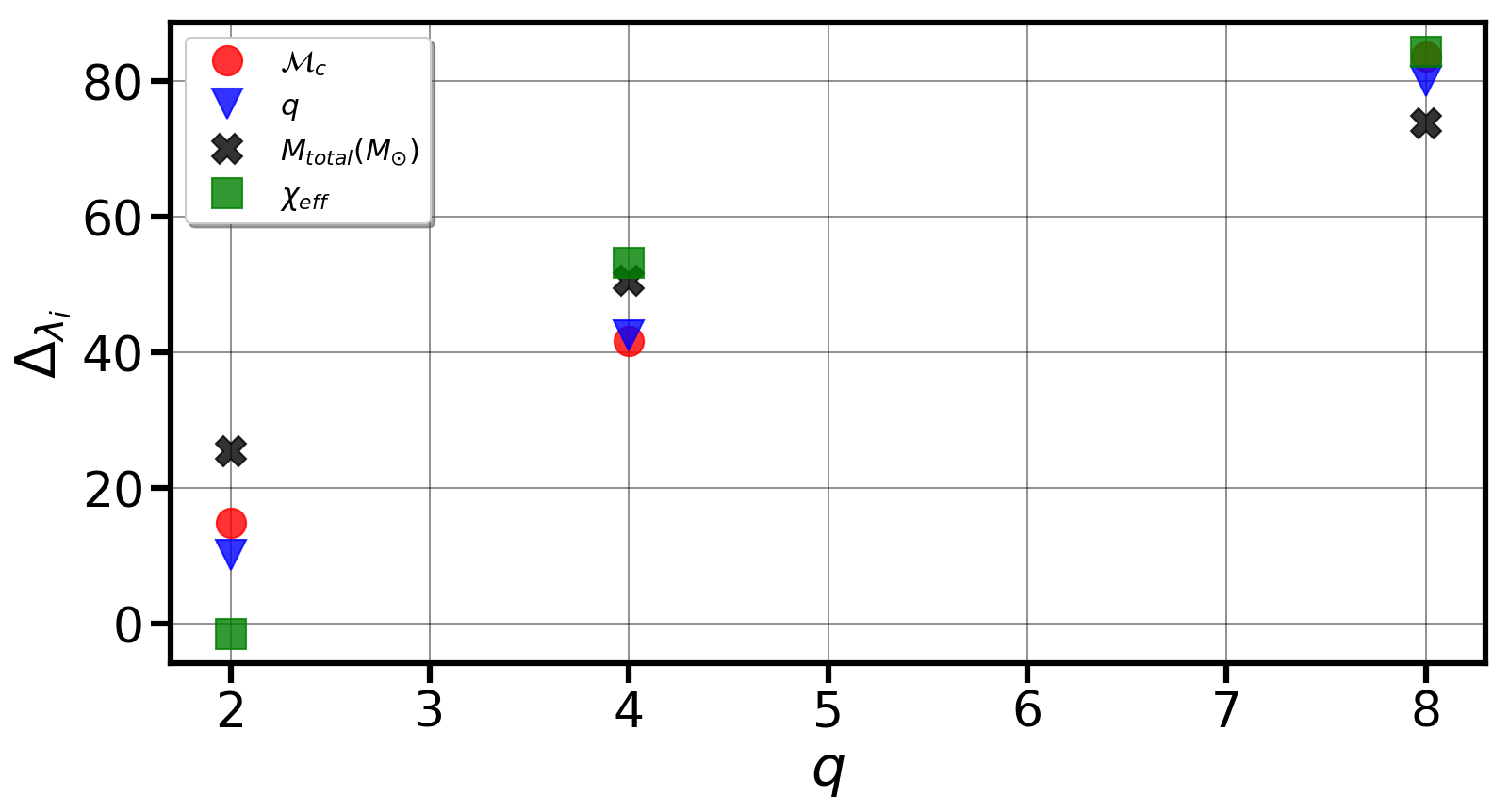}
	\caption{Plot of $\Delta_{\lambda_i}$ for all the \phenomhm\ injections. Results for face-on, 60$^{\circ}$ and 90$^{\circ}$ inclination injections are shown in the top, middle and bottom panels respectively. $\Delta_{\lambda_i}$ for the parameters $(\mathcal{M}_{c}, M_{total}(M_{\odot}), q, \chi_{\rm eff})$ are shown with red-circle, black-cross, blue-lower triangle and green-square respectively. }~\label{fig:phnhm_constraints}
	\end{figure}

		For a given (mass-ratio, inclination) configuration, the improved constraints on the inclination and distance parameters by using multi-mode templates might lead to better constraints on the intrinsic parameters. To check for that, let us define $d^{model}_{\lambda_i} = C^{upper}_{\lambda_i} - C^{lower}_{\lambda_i}$, where $C^{upper}_{\lambda_i}$ and $C^{lower}_{\lambda_i}$ are the upper and lower bounds of the 90\% CI for a given parameter $\lambda_i$. Hence, $d^{model}_{\lambda_i}$ would provide a measure of the posterior width. Using this, we define the relative percentage difference between the credible interval widths for a given configuration ($\Delta_{\lambda_i}$) as,
	\begin{equation}
	\Delta_{\lambda_i} = 100 \left( \frac{d^{\phenomd}_{\lambda_i} - d^{\phenomhm}_{\lambda_i}}{d^{\phenomd}_{\lambda_i}} \right).
	\end{equation}
	For a given intrinsic parameter, $\Delta_{\lambda_i}$ would quantify the improvements on the parameter constraints from using multi-mode templates.

For the following plots,  the posterior over a parameter from each run is clipped within its 90\% credible intervals and we plot
them as a violin plot.
For each parameter, the y-axis shows the value of the recovered posterior and the x-axis gives the injected inclination-recovery waveform combination.
For example, if the recovery is for an edge-on injection by \phenomd, it is labelled as $\iota = 90^{\circ} \text{PhnD}$.
Posteriors for $q =2, 4, 8$ systems are shown in Blue, Grey and Orange respectively.  We will first discuss the results of
\phenomhm\ injections.

\subsubsection{ \phenomhm\ Injections}~\label{sec:phnhm_intrinsic_recov}

The results of recovered intrinsic parameters for the \phenomhm\ injections are given in Fig.~\ref{fig:PhenomHM_intrinsic}.
The percentage improvement in parameter measurements, $\Delta_{\lambda_i}$, is shown for
$\lambda_i \in (\mathcal{M}_c, M_{\rm total}, q, \chi_{\rm eff})$ in Fig.~\ref{fig:phnhm_constraints}.

We consider the face-on systems first, which are the left-hand columns in each of the panels in Fig.~\ref{fig:PhenomHM_intrinsic}, and the upper panel in Fig.~\ref{fig:phnhm_constraints}. For all mass ratios, at face-on inclination the posteriors recovered by \phenomhm\ and \phenomd\ are consistent with each other, show no bias and are recovered at very similar maximum likelihood. This is expected due to the \emph{almost} zero contributions ($<10\%$) of higher-order modes to total signal power at face-on inclination and due to the underlying quadrupole model for \phenomhm\ being \phenomd.

The log Bayes factor between \phenomhm\ and \phenomd\ $\log(B^{PhnHM}_{PhnD})$ for all face-on systems is $\leq 4$. At $q=2$, the confidence intervals are almost identical between
\phenomhm\ and \phenomd\ recovery, although the mass-ratio recovery shows a greater preference for lower $q$. Any improvements in the measurement precision are difficult to detect in the posterior plots, but are clear in Fig.~\ref{fig:phnhm_constraints}.
We see that $q$ is always recovered slightly better by \phenomhm\, even for the $q=2$ system, with an improvement of $\sim$5\%, with the
recovery of the other parameters slightly \emph{less} constrained at $q=2$.  Although the effect is small, for face-on signals, the higher-mode content \emph{does} lead to a slight improvement at higher mass ratios, where we see that
the total mass and effective spin are recovered more accurately, and the effective spin measurement is also more precise.

\begin{figure*}

\includegraphics[width=1.\textwidth]{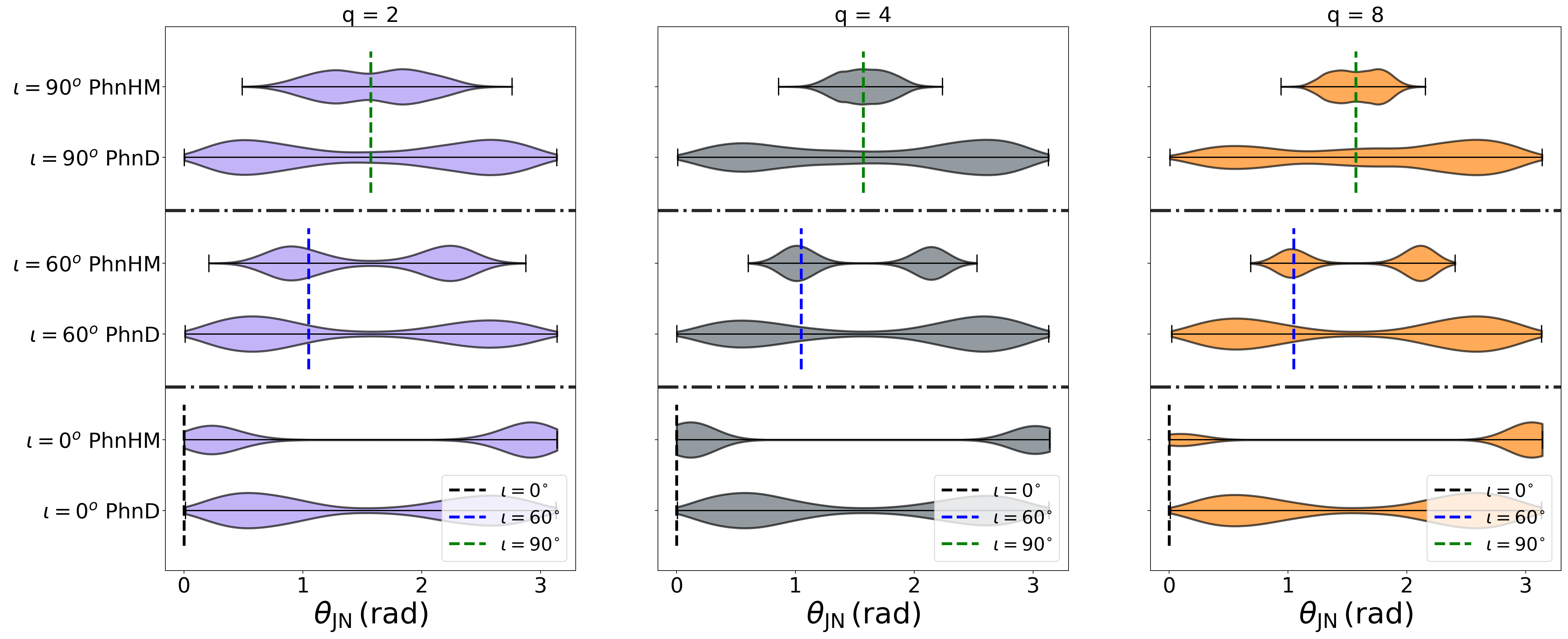} \\
\includegraphics[width=1.\textwidth]{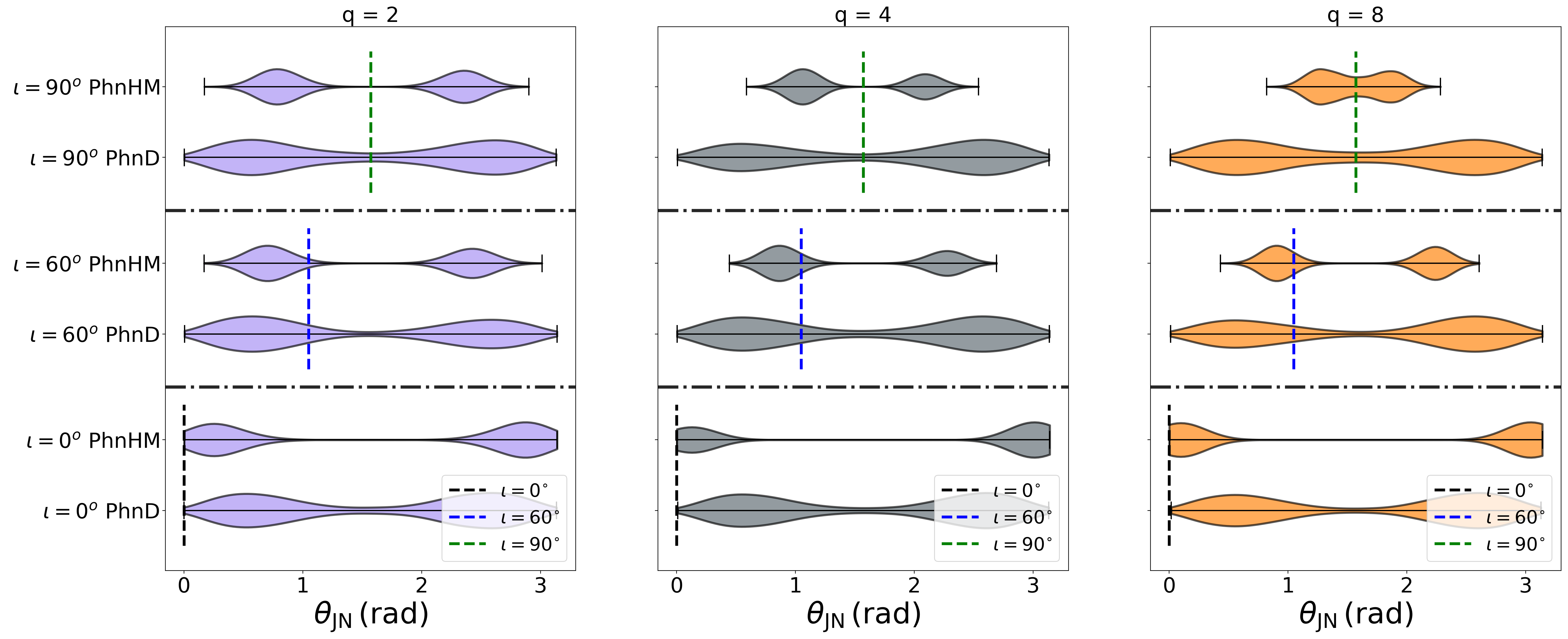}

\caption{  Recovery of $\theta_{JN}$ for \phenomhm\ injection (top row) and hybrid NR injections (bottom row) for inclinations 0,
$60^{\circ}$  \& $90^{\circ}$  and with \phenomhm\ and \phenomd\ as recovery waveform models. Inclination recovery for $q =2, 4, 8$
configurations are shown in the left, centre and right columns respectively, with the recovery for each inclination separated by horizontal
dashed-black lines. The true value of the injections are given in dashed Black, Blue and Red lines for 0, $60^{\circ}$  \& $90^{\circ}$.
}
\label{fig:theta_jn_recovery}
\end{figure*}

Now consider systems with inclination 60$\degree$. The effects of higher modes are perhaps the most relevant for these cases, because this is where we statistically expect to observe greater number of signals; this is clear from, for example, the \phenomd\ inclination recoveries in Fig.~\ref{fig:theta_jn_recovery}, which
predominantly recover the observational prior.
In the middle column of each Fig.~\ref{fig:PhenomHM_intrinsic} panel, we see that the \phenomd\ recovery starts to show a bias away
from the true values and the parameters are recovered at comparatively lower likelihood.
These effects are stronger with increasing sub-dominant mode contribution to total signal power.
At $q=2$, the \phenomd\ recovery of $\mathcal{M}_c$
and $M_{\rm total}$ are slightly biased away from the true value towards overall higher total mass and more equal mass.
For a given $q$, the waveform length decreases (increases) at higher (lower) total mass or more negative (positive) $\chi_{\rm eff}$.
For the $q=2$ system, the shift to an overall higher total mass is compensated by higher $\chi_{\rm eff}$ recovery.
We might expect that the
 increase in total mass puts more power into the signal, to mimic the extra power that is there from the higher harmonics,
 but we see the opposite trend for mass ratios $q=4$ and $q=8$. Regardless of the parameter shifts in the quadrupole-only
 model to find the best agreement with the
 higher-mode signal, it is clear that best matching \phenomd\ signals does not agree especially well with the true signal, as indicated by the
 drop in log-likelihood for signals with increasing mass ratio.

Biases in \phenomd\ recovery increase with inclination and are most extreme for edge-on cases. Note that there are also biases in the
\phenomhm\ recovery, but these are caused by the prior. The inclination prior has very low support from edge-on inclinations, and hence,
the recovered $\theta_{JN}$ posterior tends to have more support from non-edge-on inclinations. This leads to an overestimation of the distance.
The amplitude ($\mathcal{A}$) of a BBH source is $\mathcal{A} \propto \mathcal{M}_c^{5/6}/d_{L} = M^{5/6} \sqrt{\eta}/d_{L}$. At higher masses,
$\mathcal{M}_c$ and $M_{\rm total}$ are the better constrained mass parameters. Hence, overestimating $d_{L}$ (with good constraints on
$\mathcal{M}_c$ and $M_{\rm total}$) would lead to a higher value of $\eta$ or, equivalently, a lower $q$. This effect is what causes the slight
bias on the \phenomhm\ recovered $q$ for edge-on $q=4$ and $q=8$ \phenomhm\ injections.

Returning to the \phenomd\ recovery biases, the most extreme case we see is that of the \phenomd\ recovery of the $q=8$ edge-on
\phenomhm\ injection. The recovered posteriors show a bi-modal distribution. For this injection, \phenomd\ sees the signal as two
completely different systems with parameters $[ (M_{\rm total}, q, \chi_{\rm eff}) \sim (85, 7, -0.25), (63, 11, -1.) ]$, with comparable
(but overall very low) likelihood. Including the prior difference, the posterior values around those two areas in the parameter space are very similar, leading to the bimodality. As an additional test, we checked that the bi-modal distribution is not due to sampling error by performing two additional
nested-sampling PE runs with 2048 and 4096 live points and another MCMC run with 16 parallel chains and 5000 effective samples,
but the bi-modal distribution persists. Two additional PE runs were then done where the sky-position of the signal was randomized while
keeping the polarisation fixed and vice versa. The bi-modality seen by \phenomd\ for the the run in Fig.~\ref{fig:PhenomHM_intrinsic} is lost
for these runs, but the corresponding parameters recovered by \phenomd\ were 1) highly biased  and 2) recovered with similar low
maximum likelihood ($\Delta Log(\mathcal{L}) \sim$ -95).
The bi-modality of recovered parameters in Fig.~\ref{fig:PhenomHM_intrinsic} is a consequence of \phenomd\ seeing the signal as from
two different but equally likely systems, which is then lost when the signal morphology changes with changing sky-position and polarisation
values. But, for all sky-position and polarisation combinations, parameters recovered by \phenomd\ for $q=8$ show a consistent bias towards
lower total mass and negative $\chi_{\rm eff}$. Also, $\log(B^{PhnHM}_{PhnD})$ for
the bi-modal run is $\approx$ 94, which implies that the signal as seen by \phenomd\ is highly unlikely as compared to \phenomhm. All this suggests
that the observed bi-modality is a combined effect of the priors over the physical parameters and the inaccuracy of \phenomd\ towards
recreating the true multi-mode signal.

$\mathcal{M}_c$ posteriors recovered by \phenomhm\ are accurate for all the cases. At face-on $q=2$, recovered $q$ has large support from near-equal mass systems, but this behaviour is lost at higher inclinations.  At edge-on $q=4$ and $q=8$, mass-ratio and $M_{\rm total}$ are slightly biased towards lower values, with accurate recovery of the mass-spin parameters for all other cases.

Where quadrupole models tend to recover a biased $\chi_{\rm eff}$ at higher inclinations, \phenomhm\ recovery does not.
For the $q=2$ injections with $\theta_{JN} =$ 60$^{\circ}$ and 90$^{\circ}$, the recovered $\chi_{\rm eff}$ posteriors have almost the same width for \phenomd\
and \phenomhm\ templates ($\Delta_{\chi_{\rm eff}} \sim 0$). Although the spread of these posteriors is similar, $\chi_{\rm eff}$ recovery with \phenomhm\ is accurate, whereas \phenomd\ recovery is biased.

We now consider the relative improvement in parameter precision (i.e., the widths of the posteriors, irrespective of any bias from the true
injected values), as shown in Fig.~\ref{fig:phnhm_constraints}.

At inclinations of 60$^{\circ}$ and 90$^{\circ}$, the mass parameters recovered by \phenomhm\ are always better constrained than corresponding \phenomd\ recoveries (see middle
and bottom panels of Fig.~\ref{fig:phnhm_constraints}), i.e., $\Delta_{\lambda_i} > 0$. For a given inclination-parameter combination,
$\Delta_{\lambda_i}$ increases with increasing $q$. For e.g., for $\theta_{JN} = $ 60$^{\circ}$,
$\Delta_{\mathcal{M}_c} \sim 20\%, 30\%, 50\%$ for $q=2$, 4 and 8 respectively. The comparatively high $\Delta_{\lambda_i}$ values for the
edge-on $q=8$ configuration is due to the bi-modality of the \phenomd\ recovered posteriors. At $q=4$ and $q=8$, $\Delta_{\chi_{\rm eff}} \geq 0$ for all inclinations,
and \phenomd\ recovers biased $\chi_{\rm eff}$ posteriors for non-face-on inclinations whereas \phenomhm\ recovery is accurate for
all configurations. Overall, we observe better constraints on the
mass parameters for inclined system across the mass-ratio space.

The results in Fig.~\ref{fig:phnhm_constraints} illustrate that in addition to more accurate parameter recovery when using a multi-mode model,
we also find improved precision in the parameter measurements.
We see that other than for face-on configurations, the recovered mass and spin parameters are better constrained ($\Delta_{\lambda_i} > 0$),
with the constraints improving with increasing mass-ratio and/or inclination.

\subsubsection{ Hybrid-NR Injections}

We now consider injections of the same physical systems, but using hybrid-NR waveforms instead of \phenomhm. The purpose of this is
to assess systematic errors in the \phenomhm\ model. If the hybrid-NR waveforms and the corresponding \phenomhm\ waveforms were
almost identical, then the results from hybrid-NR injections would be nearly identical to those in the previous section. This would require
not only that the \phenomhm\ model accurately capture all of the features of the NR waveforms, but that the
numerical errors in the higher-mode content of the NR waveforms be insignificant, along with the differences between the hybrids' EOB inspiral
and the \phenomhm\ inspiral. We do not expect any of these requirements to hold, and so with hybrid-NR injections we can determine which
of the previous results is robust against uncertainties in the \phenomhm\ model, and for which parameters and regions of the parameter
space systematic errors may contaminate measurements.

 The results of recovered intrinsic parameters for Hybrid NR waveform injections are given in Fig.~\ref{fig:NR_intrinsic}. For these injections,
 we will not plot the $\Delta \lambda_{i}$s, but will discuss them when relevant.  For $q=2$, $q=4$ and $q=8$, at face-on inclinations,
 posteriors recovered by \phenomhm\ and \phenomd\ follow the same behaviour that we saw for \phenomhm\ injections. Parameters
 recovered by both models are accurate, but the posteriors recovered by \phenomhm\ show slightly improved constraints on the mass and
 spin parameters ($\Delta_{\lambda_i} \geq 0$). Given the weak higher-mode content in face-on injections, it is not surprising that the
 results do not depend strongly on whether \phenomhm\ or hybrid-NR waveforms were injected.

The same is true for $q=2$ injections at all inclinations. In some cases the posteriors are wider for the hybrid-NR injections, the most notable
example being the \phenomhm\ recovery of $M_{\rm total}$ at $\theta_{JN} = $ 60$\degree$. However, it is not too surprising that recovery is more
accurate and precise when the injection was an instance of the waveform model used for recovery, as with the \phenomhm\ injections.
The lack of any notable impact of systematic errors at $q=2$ means that the bias in the quadrupole-only recovery of $\chi_{\rm eff}$
for the 60$\degree$ injection is a robust result.

For $q=4$ and $q=8$ injections, we do see some clear differences between the results from \phenomhm\ and hybrid-NR injections.
To ease the interpretation of a large number of results, we first focus on the the recovery using \phenomhm, which is most relevant to future
GW observations. Here we see that for most parameters the recovery again does not change significantly between the two classes of
injections. There are two exceptions. One is the recovery of the mass ratio for $q=8$ edge-on systems, where we see that the prior effect,
which leads to an underestimation of $q$, becomes yet more pronounced with the hybrid-NR injection, and the true value is outside the 90\% C.I.
The other is the measurement of the total mass. The posteriors are broader in the $q=2$ case for both 60$^{\circ}$ and edge-on configurations,
and for $q=4$ and $q=8$ the mass is clearly biased. However, we also note that these are cases where the quadrupole-only \phenomd\
shows extremely large biases, and the \phenomhm\ recovery shows a clear improvement.

As with the biases on mass parameters, biases on \phenomd\ recovered $\chi_{\rm eff}$ for the hybrid-NR injections follow the same behaviour as for \phenomhm\ injections for $q =$ 2 and 4, but the bias for $q = 8$ is in the opposite direction. However, \phenomhm\ is able to measure the true value of $\chi_{\rm eff}$ within its 90\% CIs for all the configurations.

If we now look at the quadrupole-only recovery, we see that there are many differences between the \phenomhm\ and hybrid-NR injections.
Some of these are counter-intuitive: for example, for the $q=8$ - 60$^{\circ}$ injections, the bias in the chirp mass recovery is in opposite
directions for the two classes of injections. However, we note that in all of these examples, the log-likelihood for the \phenomd\ recovery is
low, and so it is possible for the parameters of best agreement between \phenomd\ and the injected waveform to show greater variation;
and note also that these will also be sensitive to all of the intrinsic and extrinsic parameters.

Overall we conclude that only in the most extreme cases ($q=8$ and edge-on) do we see a risk of biases when using \phenomhm\ for
parameter recovery, and in all cases it shows an improvement, and often a dramatic improvement, over a quadrupole-only model.

\section{Recovery of extrinsic parameters}
\label{sec:extrinsic}

We saw in Ref.~\cite{London2018} that one of the most significant impacts of using a higher-multipole model for parameter measurement is in
the recovery of the binary's inclination ($\theta_{JN}$) and distance ($d_L$). In a quadrupole-only model, the only effect of changing the
inclination is to change the overall amplitude, which is degenerate with a change in distance. The strength of the $(2,2)$ mode varies by a
factor of two between face-on and edge-on systems (for the plus polarisation), leading to an uncertainty of roughly a factor of two
in the distance measurement.
The inclination measurement is then dominated by the prior, which is a combination of the inclination dependence of the detector sensitivity
(the sensitivity is twice as sensitive to face-on systems) and the inclination probability distribution, which is uniform in $\cos(\theta_{JN})$.
The result is a distribution that peaks at $\sim$30$\degree$ and $\sim$150$\degree$, and this is reflected in the
 \phenomd\ inclination recovery
plots below. In general, as was seen in Ref.~\cite{Graff:2015bba, Littenberg2012, OShaughnessy:2013zfw, OShaughnessy:2014shr}, the use of higher-order mode templates break the degeneracy present between
$\theta_{JN}$, $\psi$  and $\phi$ that allows for better measurements of inclination value and hence better distance precision. We see that \phenomhm\ is able to capture inclination information better than \phenomd\ and leads to improved constraints on the distance.

In the following sections we quantify these effects for both \phenomhm\ and hybrid-NR injections;  the former quantify the impact
of higher multipoles, while the latter allow us to investigate systematic biases due to approximations in \phenomhm.

\begin{figure*}

\includegraphics[width=1.\textwidth]{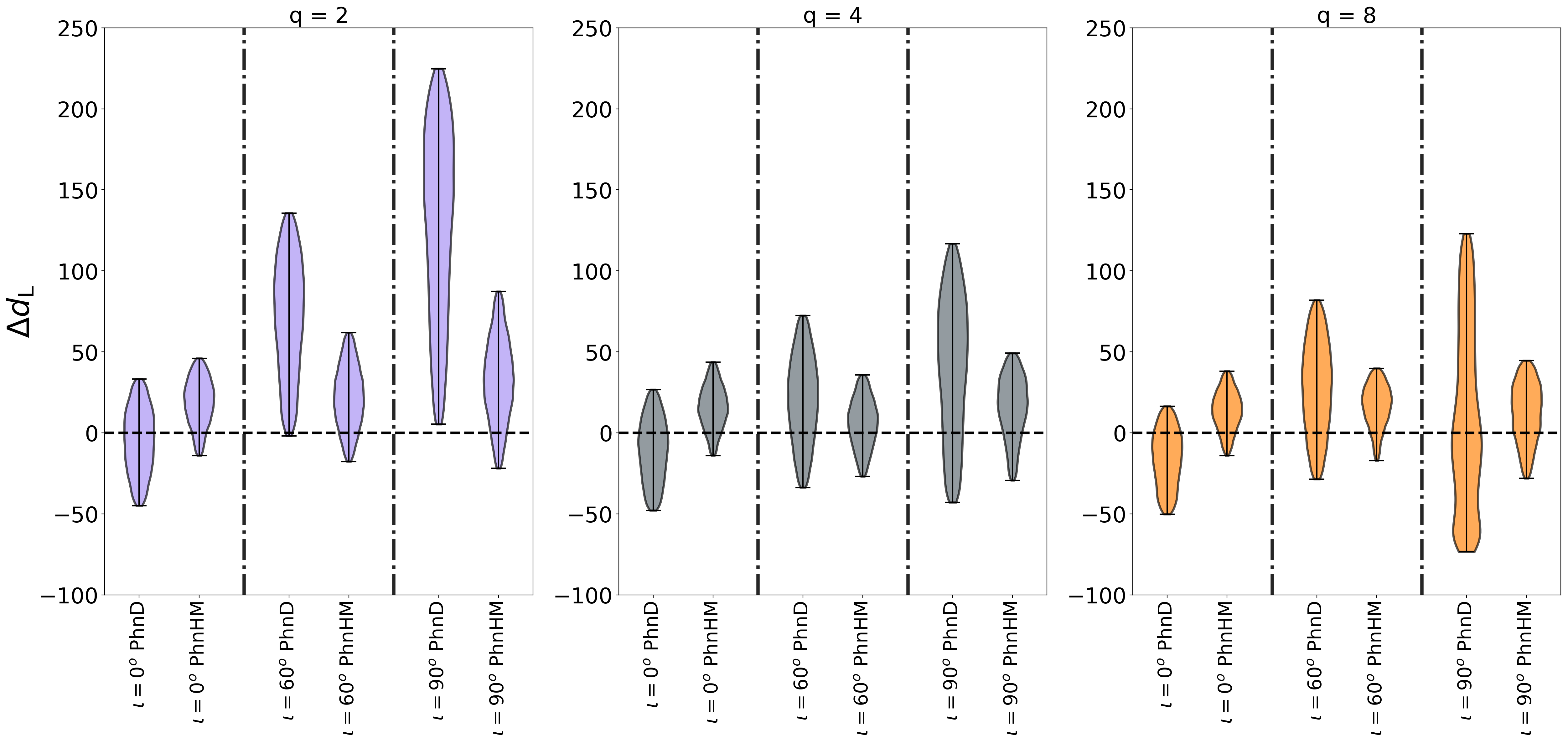} \\
\includegraphics[width=1.\textwidth]{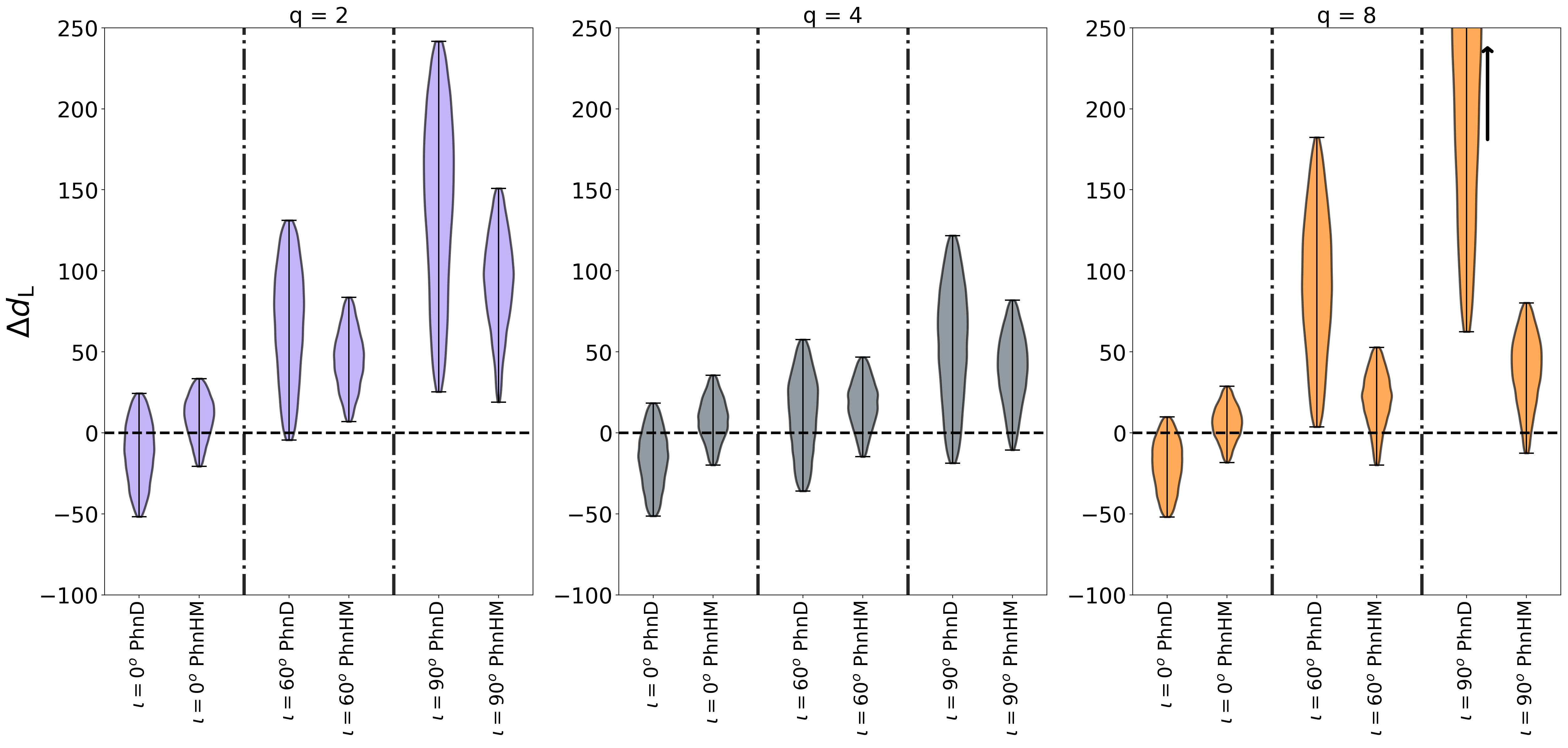}

\caption{ Recovered percent distance error $\Delta d_{L}$ (see Eq:\ref{eq:dist_recov_error}) recovery for \phenomhm\ injection (top row) and Hybrid NR injections (bottom row) for inclinations 0, 60$^{\circ}$ and 90$^{\circ}$ with \phenomhm\ and \phenomd\ as recovery waveform models. Distance recovery for $q = 2, 4, 8$ configurations are shown in the left, center and right columns respectively. The $\Delta d_{L}$ = 0 line is shown in horizontal dashed-black line, with the vertical dashed lines separating recovery for different inclinations. The injected distance value for \phenomhm\ [NR] injection for $q= (2, 4, 8)$  0$\degree$ is (895, 624, 388) [880, 639, 398] Mpc,
60$\degree$ is (537, 404, 258) [523,  376,  249] Mpc and for 90$\degree$ is (387, 307, 199) [367, 253, 183] Mpc. For the $q=8$,
90$\degree$ hybrid-NR recovery by \phenomd, $\Delta d_{L}$ extends up to 400\%. }
\label{fig:distance_recovery}
\end{figure*}

\subsubsection{$\theta_{JN}$ recovery}

Fig.~\ref{fig:theta_jn_recovery} shows the results for inclination recovery for both the \phenomhm\ and hybrid-NR injections.
For both \phenomhm\ and hybrid NR injections, at all mass-ratio and inclination configurations, $\theta_{JN}$
recovery by \phenomd\ shows a
similar bimodal behaviour and mostly follows the prior, as discussed above. Inclination recovery is unaffected by mass-ratio
or inclination value
for the quadrupole only model and thus it is not possible to differentiate between a non-inclined and inclined system.

For \phenomhm\ recovery, the bi-modality for the inclination posterior persists, but the posteriors are better constrained. At face-on
configurations for \phenomhm\ and hybrid NR injections, \phenomhm\ sees the system as strongly face-on or face-off.
For 60$^{\circ}$ \phenomhm\ injections, the recovered inclination is peaked near the true value and the constraint on the inclination improves
with increasing mass-ratio. The edge-on \phenomhm\ injection posteriors show a similar behaviour. For Hybrid NR injections, inclination recovery
for 60$^{\circ}$ is peaked just off the true value and for edge-on $q=(2,4)$ systems, the recovery is strongly biased.
This is not surprising: the systematic errors in
the \phenomhm\ model enter almost entirely into the higher multipoles, and so the largest systematic errors will be observed for edge-on systems,
where the higher multipoles contribute most to the signal. We find, however, that it is \emph{only} for the edge-on cases that this bias appears.
Since edge-on systems are still roughly half as strong as equivalent face-on systems, they are eight times less likely to be observed.

\subsubsection{Distance recovery}

Recall that injections were made such that the signal's SNR was 25 in all cases. Since the signal strength decreases
as mass-ratio increases,
and also as the inclination varies from face-on to edge-on, the edge-on $q=8$ system is injected at a much smaller
distance (199\,Mpc) than the
face-on $q=2$ system (895\,Mpc). Although there is some variation in the injection distance between the \phenomhm\
and hybrid-NR injections,
due to the differences in their higher-multipole content; these are small and are always less than 10\%. All of the injection distances are given in
the caption to Fig.~\ref{fig:distance_recovery}.

Fig.~\ref{fig:distance_recovery} shows the results for distance recovery. We plot the relative percent distance error, which we define as,
\begin{equation}~\label{eq:dist_recov_error}
\Delta d_{L} = 100 \left( \frac{d_{L}^{\rm posterior} - d_{L}^{\rm injected}}{d_{L}^{\rm injected}} \right)
\end{equation}
For \phenomhm\ injections, the true distance value lies within the 90\% confidence intervals for most of \phenomd\ and all of \phenomhm\ recovered posteriors. At larger inclinations, the quadrupole model tends to overestimate the distance to the binary. For $q=2$ Hybrid NR injections,
at inclinations 60$\degree$ and 90$\degree$, 90\% CIs for $d_{L}$ recovered by \phenomhm\  do not include the true value. For all other situations though, 90\% CIs for $d_{L}$ recovered by \phenomhm\ contain the true injected value.

	For quadrupole-only templates, as the recovered inclination is the same for all injected inclinations, the recovered distance for non-zero inclinations tend towards overall larger values leading to larger $\Delta d_{L}$. For \phenomhm\ injection - \phenomhm\ recovery, where injected $\theta_{JN}$ lies within the 90\% CIs of recovered $\theta_{JN}$, the real distance is recovered at all times. For the $q=2$ and $q=4$ hybrid-NR injection - \phenomhm\ template, recovered $\theta_{JN}$ at 60$^{\circ}$ (90$^{\circ}$) is slightly (completely) off the true value which causes the recovered distance to be overestimated from the true value. This is likely due to the different mode content in the signal and template and the waveform inaccuracies in \phenomhm. But, these results do indicate that use of multi-mode template waveform will lead to better distance measurements.

The improved constraints on inclination for \phenomhm\  recovery translates to improved constraints on the measured distance of the
system as compared to \phenomd\ recovery.  We see this behaviour for all the configurations. At face-on configurations, \phenomhm\ constrains
the distance about $\sim$ 20\% - 25\% better as compared to \phenomd. For higher inclinations, the constraint improves by about
$\sim$ 30\% - 60\%.

			\begin{figure*}
	  \includegraphics[width=0.95\linewidth]{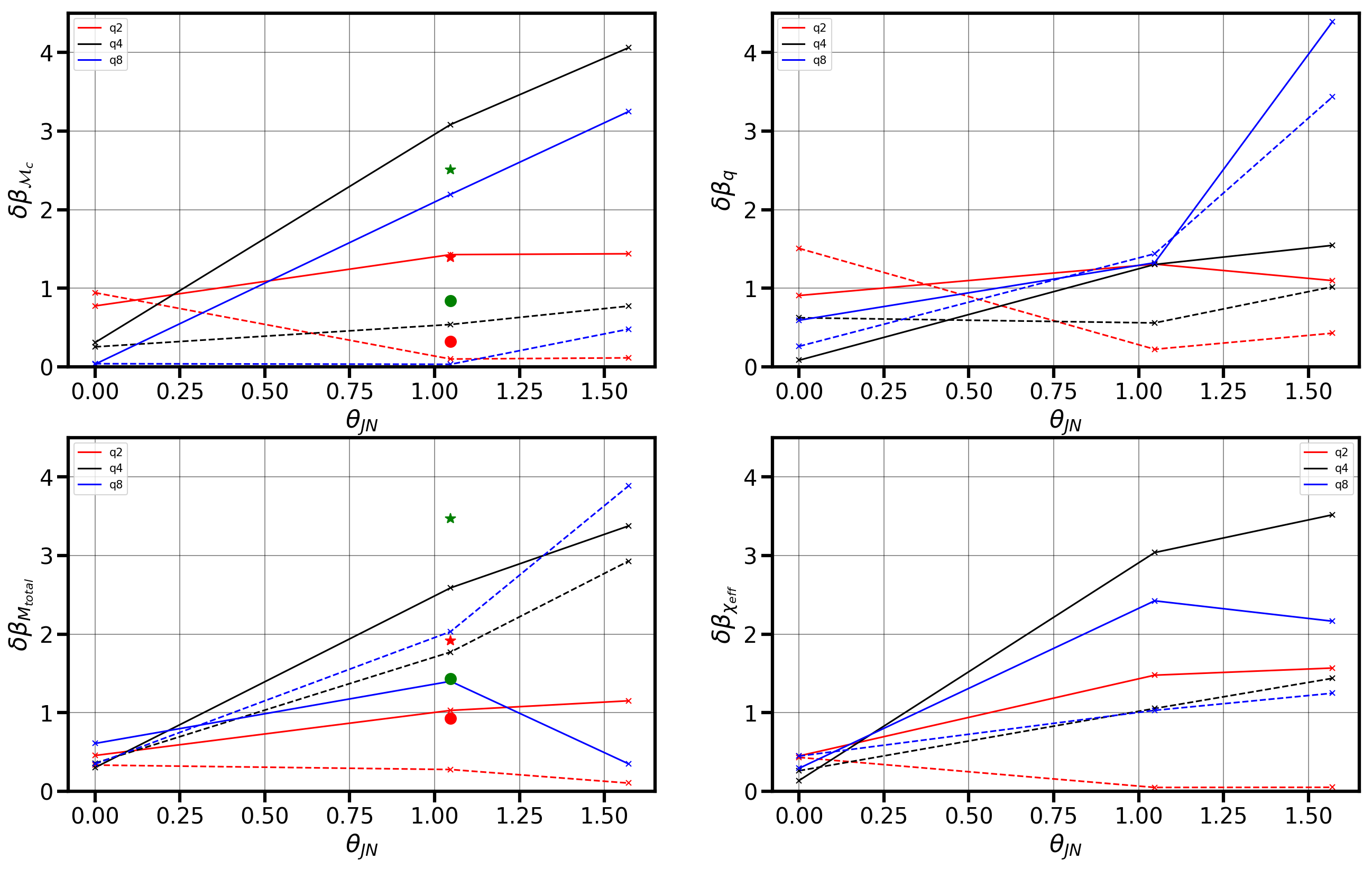}
	\caption{We plot the quantity $\delta \beta_{\lambda_i}$ for the parameters $(\mathcal{M}_c , M_{total}, q, \chi_{\rm eff})$ for
hybrid-NR injection results with the solid (dashed) lines indicating the bias value for \phenomd\ (\phenomhm) recovery. $\delta \beta_{\lambda_i}$ for $q=2, 4 8$ are shown in Red, Black and Blue respectively. The systematic bias for the $q=2, 6$,
$M/M_\odot = 51, 56$ and SNR = 48 configurations from Ref.~\cite{Littenberg2012} are shown in red (green) with the quadrupole
(multi-mode) recovered bias shown with a star (circle). }~\label{fig:nr_injs_bias_all}
\end{figure*}

\section{Conclusions}

This is the first study that quantifies the accuracy of inferred source parameters using a multi-mode aligned-spin model
waveform. To do that, we consider two families of non-spinning multi-mode signal waveforms (\phenomhm\ and hybrid-NR) over a range of mass-ratios
($q = 2,4,8)$ and inclinations (face-on, 60$\degree$, and edge-on), with fixed total mass, and compare the parameters recovered
by multi-mode and quadrupole-only templates. In all cases we consider a signal-to-noise ratio (SNR) of 25.
We fix the total mass of injected signals at 100$M_{\odot}$, for three reasons; i) the relative measurable signal power in the higher
modes increases with mass, and so the choice of a high mass allows us to provide an estimate of the largest impact of higher modes
in likely LIGO-Virgo observations, for which total binary masses above 100$M_\odot$ will be rare. ii) \phenomhm\ is an approximate
waveform model of the sub-dominant modes that is not tuned to NR waveforms, and the most uncertain part of the \phenomhm\
modelling is in the merger and ringdown phases. Hence, the choice of a high total mass also allows us to also make a conservative
estimate of the systematic errors due to waveform inaccuracies. iii) \phenomhm\ is a more computationally expensive model than its
quadrupole-only counterpart, and signals with total mass 100$M_{\odot}$ allow us to quickly perform a large PE study. Although, as previous studies have shown~\cite{PhysRevD.93.084019, PhysRevD.96.124024, Varma:2014jxa}, systematic errors due to neglecting higher-order modes in template waveforms increase at higher masses. Optimized versions of the model will make it easier to perform a much more extensive study over a wider range of parameters,
including much lower masses.

Here is a summary of our main results.

Our key results on measurements of intrinsic parameters are in Figs.~\ref{fig:PhenomHM_intrinsic} and \ref{fig:NR_intrinsic}.
For face-on systems there is no appreciable bias in the quadrupole-only recovery at any mass ratio. There are biases in the total
mass of up to 10\% at 60$\degree$ inclination, and 20\% or 30\% for edge-on configurations. For the edge-on $q=8$
configuration (\phenomhm\ injection), the signal matches the quadrupole-only model so poorly that the recovery is bimodal. Mass ratio is strongly biased
toward equal-mass recovery for small mass ratios, but shows less bias at higher mass ratios.
(This point is relevant to the observation GW170729~\cite{Chatziioannou:2019dsz}, where the templates with higher-order modes
were able to resolve $q$ between 1.25 and 3.3 at 90\% confidence, i.e., providing strong evidence that the mass-ratio was bounded
away from equal mass, while quadrupole-only models gave a 90\% C.I. from $q=1$ up to $q = 2.5$.)
 The effective spin $\chi_{\rm eff}$
shows a bias of up to 0.2 for 60$\degree$ inclination and cannot be measured at all in some edge-on cases. Our results suggest
that bi-modal parameters or, in less extreme cases, double-peaked posteriors, might occur (for some cases) when measuring high-mass-ratio
systems with a quadrupole-only model which could be resolved with multi-mode models. The overall lower likelihood of recovered parameters by the quadrupole-only model (compared to multi-mode model) at high-mass ratio - high-inclination combination suggest that the model cannot be trusted for accuracy in that region.

For \phenomhm\ injections, recovery with a higher-mode model not only removes these biases (as we would expect, since the injection and recovery use the same
model), but also increases the precision of the measurement. Fig.~\ref{fig:phnhm_constraints} shows the percentage improvement
in the size of the 90\% credible intervals over using a quadrupole-only model. The improvement is up to 50\% for 60$\degree$
inclination and up to 80\% for edge-on configurations; the improvement is roughly linear with mass ratio~\footnote{Note that improvement of ~80\% for $q=8$ edge-on is due to the bimodal recovery of quadrupole-only model}. Improvement in
parameter recovery was also considered in Ref.~\cite{Graff:2015bba}, but using a higher-mode model that did not include
spin. We find that the addition of the spin dimension can significantly increase the widths of the confidence intervals. For example,
recovering a $q=4$, $M = 100\,M_\odot$, 60$\degree$-inclination SNR=18 signal with a nonspinnng model leads to an uncertainty
in the chirp mass of $\Delta \mathcal{M}_{obs}/\mathcal{M}_{obs} = 0.056$~\cite{Graff:2015bba}. If we naively rescale to an SNR of 25,
the uncertainty would decrease to $\sim 0.04$. By contrast, the uncertainty when using a spinning higher-mode model is 0.168,
i.e., four times larger.

NR-hybrid injections show broadly consistent results, indicating that the systematic errors in the \phenomhm\ model are in general
small. The \phenomhm\ recovery gives comparable results between the \phenomhm\ and NR-hybrid injections, except for biases
of up to 5\% in the total mass for $q=4$ and $q=8$ non-face-on cases; this is less than the bias
incurred by using a quadrupole-only model. Results show larger differences in the \phenomd\ recovery between the two injection
sets, but these are cases where the log-likelihood is poorer, and so we ascribe these less significance. Our conclusion is that the
\phenomhm\ model leads to improved parameter measurements over a quadrupole-only model in \emph{all} cases; and
except for high-mass high-mass-ratio high-inclination signals with an SNR of 25 or higher, systematic errors will not affect results.

This is quantified further in Fig.~\ref{fig:nr_injs_bias_all}. Here we follow Ref.~\cite{Littenberg2012}, and plot the ratio of the
systematic error to the statistical error, $\delta \beta_{\lambda^i}$, for the parameter $\lambda^i$.
By ``systematic error'' we mean the difference between the true injected parameter and the
mean of the marginalised 1D recovered posterior (in Ref.~\cite{Littenberg2012}, the authors use the difference between the maximum a posteriori value and injected value), and by ``statistical error'' we mean the standard deviation of the
 posterior. Since the standard deviation corresponds to the 68\% C.I., $\delta \beta_{\lambda^i}$ is a more conservative estimate
 than if we had used the 90\% C.I. As such, if the ratio is below $\sim1.645$, then the systematic error is within the
 90\% C.I. and the measurement is not considered to be biased.
 Fig.~\ref{fig:nr_injs_bias_all} shows this error ratio for the chirp mass, total mass, mass ratio,
 and effective spin. Also shown are the results from Ref.~\cite{Littenberg2012}, although care should be taken in comparing the
 results, since that study considered lower masses, higher SNRs and different mode content for the injections.

Inclination recovery is always improved when recovering with a higher-mode model. With quadrupole-only recovery, the
distance-inclination degeneracy means that largely the same posterior is recovered, regardless of inclination
(see Fig.~\ref{fig:theta_jn_recovery}), while the higher-mode model is able to constrain the inclination. The trend in the
\phenomd\ distance recovery is consistent with expectations: for a signal with a high non-zero inclination, a quadrupole only
template model gets more support from non-edge on inclinations, suggesting that the (comparatively) weak signal is from further away
and therefore $d_{L}$ would be overestimated. A multi-mode template model can better constrain the degeneracy between the inclination,
phase and polarisation values, leading to improved constraints on the inclination, which then translates to a better constrained
measurement of distance. Distance recovery is greatly improved with the higher-mode model. At face-on configurations, \phenomhm\ constrains the distance  about 20\% - 25\% better as compared to \phenomd. For higher inclinations, the constraint improves by
about 30\% - 60\%. With improved multi-mode models, we can expect improved inclination constrains and hence, distance
measurements, with additional improvement from those three-detector observations that have good polarisation
measurements~\cite{Aasi:2013wya, Fairhurst:2010is}.

Systematic errors in \phenomhm\ are worst for edge-on higher-mass-ratio systems, where the approximations used to produce the
higher modes are least applicable, and this shows up in the inclination recovery: we do not identify the system as so clearly
edge-on in the NR-hybrid injections compared to the \phenomhm\ injections, as seen in the top panels of Fig.~\ref{fig:theta_jn_recovery}. As it is more likely that systematics due to the model inaccuracies dominate at larger inclinations, we can expect accurate parameter recovery at lower inclinations.

As expected, priors can bias results for statistically less likely configurations, i.e., edge-on, even when using a higher-mode model.
See, for example, the edge-on recovery of the mass ratio $q$ for the $q=8$ case in Fig.~\ref{fig:PhenomHM_intrinsic}.

We note that this study is limited to binaries with total mass $M = 100\,M_\odot$, and does not include spinning signals, or precession.
The effect of black-hole spin on higher-mode contributions is much weaker than the effect of mass ratio, and given that LIGO-Virgo
observations to date suggest that astrophysical black holes in binaries predominantly have very low
spins ~\cite{LIGOScientific:2018mvr,Tiwari:2018qch}, we expect the results of this study to be relevant to the majority of observations made
with second-generation detectors. Since the impact of higher modes decreases for lower masses, the results we report here are
likely to represent the largest impact higher modes will have on GW observations.  However, studies that include spinning
configurations, lower masses, and precession, are still needed, to quantify the impact of higher modes for yet stronger signals, or
larger spins, where their effects have not yet been quantified. For this study, we had fixed the azimuth phase to zero and as was shown in~\cite{PhysRevD.93.084019}, this parameter can strongly affect parameter estimates. It would be interesting to perform a similar study, but with different azimuth phase values. Studies that consider yet higher SNRs will also benefit from more
accurate models tuned to NR simulations, although we expect that the \phenomhm\ model will be sufficiently accurate for all but the
most extreme observations.

\section{Acknowledgements}

We thank Alessandro Nagar for providing the EOB code used in generating the hybrids, and Edward
Fauchon-Jones for an early version of the hybridisation code.
We also thank Lionel London, Sebastian Khan and Frank Ohme for useful discussions.
This work was supported by Science and Technology Facilities Council (STFC) grant ST/L000962/1, European
Research Council Consolidator Grant 647839 and the National Science Foundation under Grant No. NSF PHY-1748958.
We are grateful for computational resources provided by Cardiff University, and funded by an
STFC grant supporting UK Involvement in the Operation of Advanced LIGO. This research has made use of data, software and/or web tools obtained from the Gravitational Wave Open Science Centre (https://www.gw-openscience.org), a service of LIGO Laboratory, the LIGO Scientific Collaboration and the Virgo Collaboration. LIGO is funded by the U.S. National Science Foundation. Virgo is funded by the French Centre National de Recherche Scientifique (CNRS), the Italian Istituto Nazionale della Fisica Nucleare (INFN) and the Dutch Nikhef, with contributions by Polish and Hungarian institutes

\bibliography{phenomhmpe.bib}

\end{document}